\documentclass[sigconf,authordraft,table,xcdraw,review=false]{acmart}

\AtBeginDocument{%
  }

\usepackage{subfig}
\usepackage{graphicx}
\usepackage{amsmath}
\usepackage{amsthm} 

\usepackage{amssymb}
\usepackage{multirow}
\usepackage{makecell}
\usepackage{rotating}
\usepackage[T1]{fontenc}
\usepackage{float}
\usepackage{bm}
\usepackage{amsmath}
\usepackage{amssymb}
\usepackage{amsfonts}
\usepackage{dutchcal}
\usepackage{booktabs}
\usepackage{makecell}
\usepackage{diagbox}
\usepackage{multirow}
\usepackage{bbding}
\usepackage{textcomp}
\usepackage{array}
\usepackage{caption}
\usepackage[T1]{fontenc}
\usepackage{aecompl}
\usepackage{bbding}
\usepackage{lipsum}
\usepackage{stfloats}
\usepackage{float} 
\newcommand{\ours}{H-NeiFi}

\preto{\abstractkeywords}{\nolinenumbers} 

\setcopyright{acmlicensed}
\copyrightyear{2018} 
\acmYear{2018}
\acmDOI{XXXXXXX.XXXXXXX}

\acmConference[Conference acronym 'XX]{Make sure to enter the correct
  conference title from your rights confirmation email}{June 03--05,
  2018}{Woodstock, NY}
\acmISBN{978-1-4503-XXXX-X/2018/06}

\begin{document}

\title{H-NeiFi: Non-Invasive and Consensus-Efficient Multi-Agent Opinion Guidance}


\author{Shijun Guo}
\email{24031212237@stu.xidian.edu.cn}
\affiliation{%
  \institution{Xidian University}
  \city{Xian}
  \country{China}
}

\author{Haoran Xu}
\affiliation{%
  \institution{SUN YAT-SEN UNIVERSITY}
  \city{Shenzhen}
  \country{China}
}

\author{Yaming Yang}
\affiliation{%
  \institution{Xidian University}
  \city{Xian}
  \country{China}
}

\author{Ziyu Guan}
\authornotemark[1]
\affiliation{%
  \institution{Xidian University}
  \city{Xian}
  \country{China}
}
\author{Wei Zhao}
\affiliation{%
  \institution{Xidian University}
  \city{Xian}
  \country{China}
}

\author{Xinyi Zhang}
\affiliation{%
  \institution{SUN YAT-SEN UNIVERSITY}
  \city{Shenzhen}
  \country{China}
}
\author{Yishan Song}
\affiliation{%
  \institution{The Chinese University of Hong Kong }
  \city{Hongkong}
  \country{China}
}









\begin{abstract}

The openness of social media enables the free exchange of opinions, but it also presents challenges in guiding opinion evolution towards global consensus. Existing methods often directly modify user views or enforce cross-group connections. These intrusive interventions undermine user autonomy, provoke psychological resistance, and reduce the efficiency of global consensus. Additionally, due to the lack of a long-term perspective, promoting local consensus often exacerbates divisions at the macro level. 
To address these issues, we propose the hierarchical, non-intrusive opinion guidance framework, \ours. It first establishes a two-layer dynamic model based on social roles, considering the behavioral characteristics of both experts and non-experts. Additionally, we introduce a non-intrusive neighbor filtering method that adaptively controls user communication channels. Using multi-agent reinforcement learning (MARL), we optimize information propagation paths through a long-term reward function, avoiding direct interference with user interactions. Experiments show that \ours\ increases consensus speed by $22.0\%$ to $30.7\%$ and maintains global convergence even in the absence of experts. This approach enables natural and efficient consensus guidance by protecting user interaction autonomy, offering a new paradigm for social network governance.
\end{abstract}



\maketitle

\section{Introduction}

With the rapid growth of social media, an increasing number of users exhibit diverse preferences across various topics~\cite{lu2022uncovering} and actively engage in online social interactions to exchange and update opinions~\cite{ledford2020facebook}. During this interaction process, phenomena such as consensus formation, disagreement, and polarization often emerge. A significant body of research has been conducted on classical opinion dynamics models~\cite{friedkin1990social, degroot1974reaching, dittmer2001consensus}, with many extensions made to address domain-specific applications~\cite{zhang2022opinion, kann2023repulsive}.
These studies cover various topics, including opinion diffusion~\cite{okawa2021dynamic}, opinion prediction and analysis~\cite{okawa2022predicting,monti2020learning}, and opinion maximization~\cite{liu2024community}.

\begin{figure}[!t]
\centering
\subfloat{
  \includegraphics[width=0.5\linewidth]{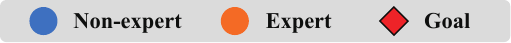}
}
\vspace{-2mm}
\setcounter{subfigure}{0}
\subfloat[Initial opinion distribution.]{
    \includegraphics[width=0.38\linewidth]{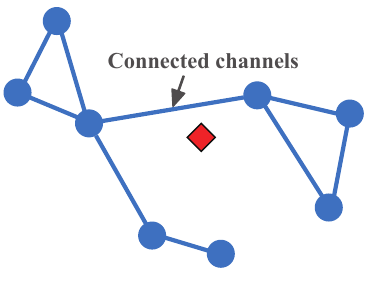}
    \label{fig:moti:init}
}
\subfloat[Echo chamber effect.]{
    \includegraphics[width=0.38\linewidth]{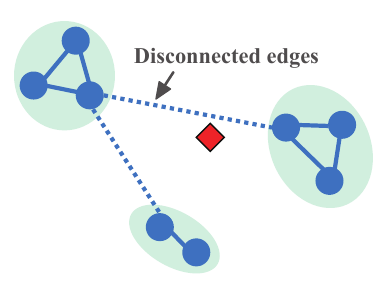}
    \label{fig:moti:echo}
}\\
\subfloat[Effect of the baselines.]{
    \includegraphics[width=0.38\linewidth]{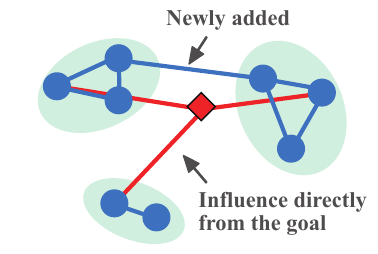}
    \label{fig:moti:base}
}
\subfloat[Effect of our \ours.]{
    \includegraphics[width=0.38\linewidth]{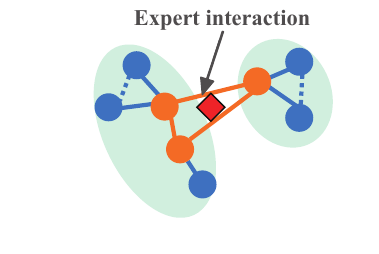}
    \label{fig:moti:ours}
}
\caption{Illustration of opinion evolution in a social network. 
(a) illustrates the initial opinion distribution, (b) demonstrates isolated clusters formed by the echo chamber effect. During the evolution process, the convergence of existing opinion baselines and our method is shown in (c) and (d). The baselines rely on direct influences, resulting in localized consensus. In contrast, our \ours\ achieves global consensus through hierarchical communication patterns and a non-intrusive neighbor filtering method.
}
\label{fig:moti}
\end{figure}

Currently, social media platforms often recommend content based on user preferences to enhance the experience and maintain engagement~\cite{gonzalez2023asymmetric}. However, this tendency exposes users to a limited range of viewpoints, reinforcing existing beliefs and leading to the echo chamber effect~\cite{cinelli2021echo}. As a result, social biases and fragmentation are exacerbated~\cite{thorson2021algorithmic,tornberg2022digital}. For example, given the initial opinion distribution shown in \figurename~\ref{fig:moti:init}, the echo chamber effect manifests in the formation of three isolated clusters, as depicted in \figurename~\ref{fig:moti:echo}. In response, recent research on opinion consensus~\cite{bernardo2024bounded,shang2022constrained,ye2020continuous} has gained traction. Some methods focus on fostering communication channels between users with significant disagreements~\cite{zhou2023sublinear,zhu2021minimizing}, while others aim to directly enhance the influence of opinions from polarized individuals~\cite{borzi2015modeling} or through goal-oriented control functions~\cite{peng2020novel,pasqualetti2014controllability}.

However, in the opinion consensus convergence process, there are several real-world deployment challenges that must be carefully considered.
Existing methods have some limitations, mainly due to two reasons: 

\textit{\textbf{Invasive}}: (i) Some approaches~\cite{zhou2023sublinear,zhu2021minimizing} add connections between users with different opinions and enforce interactions between them, while others~\cite{chen2024fully,su2021noise} introduce external opinions based on user opinion distributions to mitigate the echo chamber effect. An example of these newly imposed connections is shown in \figurename~\ref{fig:moti:base}. 
This intervention disrupts the social autonomy of user selection. 
As a result, exposure to belief-conflicting content (e.g., political values) can induce psychological reactance, paradoxically reinforcing users' original positions~\cite{reynolds2019psychological}.
(ii) Some approaches~\cite{chen2021Influence,wongkaew2015control} directly introduce influence from the opinion goal to facilitate consensus. In these approaches, users with significant opinion differences are directly controlled by a goal-oriented function, as shown in \figurename~\ref{fig:moti:base}. 
However, opinion formation is inherently a continuous evolution process based on individual cognition and social interaction. 
The discrete numerical modifications imposed by these methods lack a solid behavioral science foundation and may trigger psychological resistance mechanisms, such as the backfire effect~\cite{chen2021opinion}.

\textit{\textbf{Ineffective}}: Existing guidance methods~\cite{li2020consensus,herty2018suboptimal} often overlook that public opinion interaction constitutes a long-term, complex dynamic process.
As a result, they drive users to form stronger connections with opinion-aligned peers~\cite{musco2018minimizing}, over-optimizing local cohesion risks creating self-reinforcing echo chambers that amplify polarization and foster extreme communities at the macroscopic level~\cite{boxell2017internet}. Such localized consensus paradoxically undermines global harmony and establishes a detrimental feedback loop. Effective strategies must, therefore, balance immediate efficiency with long-term stability, avoiding self-reinforcing echo chambers while steering consensus toward socially aligned outcomes.

Based on the above analysis, we propose \ours, a novel opinion guidance framework that integrates \underline{h}ierarchical user behavior patterns with a non-invasive \underline{nei}ghbor \underline{fi}ltering approach. The goal is to guide users toward inclusive consensus on a predefined opinion goal~\cite{hegselmann2014optimal}. Users are categorized into experts and non-experts. Experts communicate exclusively among themselves, while non-experts interact with peers and assimilate opinions from experts. We model opinion dynamics using a bounded confidence model that incorporates stubbornness~\cite{xu2022effects} and attention goals.

In the non-invasive neighbor filtering approach, \ours\ introduces a pruning-based guidance paradigm: Promotion Communication Pattern (PCP) and Attention Communication Pattern (ACP). 
PCP accelerates consensus propagation by connecting low-divergence experts aligned with the global goal, while ACP innovatively incorporates a Multi-Agent Reinforcement Learning (MARL)~\cite{li2022applications} framework, modeling non-experts as agents. Leveraging the dynamic planning capabilities of MARL, ACP balances communication efficiency with long-term consensus objectives.
Notably, our method does not directly modify user opinions or inject external social nodes. Instead, it implicitly optimizes information propagation paths by dynamically adjusting the visibility of users' existing neighbors. This ``neighbor filtering'' offers two advantages: 
(i) By selectively hiding social relationships (e.g., temporarily masking non-goal-oriented neighbors), this approach feels more natural and is less likely to provoke resistance from users compared to adding edges. This is because users are generally less sensitive to the disappearance of relationships and rarely notice temporary connection losses during routine interactions~\cite{eslami2015always}.
(ii) Interventions are strictly confined to local interaction partners of users, avoiding trust costs associated with unfamiliar ones. For example, when filtering certain neighbors of a non-expert, the remaining information sources consist of weakly trusted local peers rather than unfamiliar strangers, thereby reducing the risk of cognitive conflict. 
This mechanism can make the users' social environment more harmonious and ensure the opinion guidance effect.

In social networks with multiple experts, the hierarchical guidance provided by \ours\ demonstrates an average acceleration $22.0\%$ to $30.7\%$ in consensus formation. Remarkably, it remains effective even in expert-absent scenarios, enabling unified consensus. The visual effect of \ours\ in \figurename~\ref{fig:moti:ours} further showcases our superior performance.

In summary, our main contributions are three-fold: 
\begin{itemize}
\item We propose an opinion guidance framework (\ours) based on hierarchical user interaction patterns (experts vs. non-experts), achieving goal-oriented global consensus convergence through non-invasive interventions.

\item We develop a MARL-driven dynamic neighbor filtering mechanism that adaptively adjusts local communication channels via long-term planning strategies, effectively avoiding invasive interference and reducing local opinion fragmentation.

\item Extensive experiments across various social scenarios of different scales demonstrate the effectiveness and superiority of our approach.
\end{itemize}




\section{Related Work}
In studies on opinion guidance, where users are guided to converge toward a specific goal, the common method is to influence opinion changes through external interventions. Kurz et al.~\cite{kurz2015optimal} proposed a method involving ``strategic agents'', whose opinions directly shape those of others. 
Dietrich et al.~\cite{dietrich2017control} introduced a leadership-based model with bounded confidence, where users are drawn toward a leader who then guides them to the target opinion.
Cicolani et al. \cite{cicolani2024opinion} explored how consensus can be achieved in scenarios where agents are influenced by a common influencer, even under conditions of non-universal interaction and time-delayed coupling.
In these methods, the leader's opinion is controlled and has a direct influence on the opinions of the followers.

Some methods have also integrated reinforcement learning (RL) into consensus promotion.  As a machine learning technique, RL allows agents to maximize cumulative rewards through interactions with the environment. Wang et al.~\cite{wang2020opinion} proposed an RL-based opinion management framework, where a moderator adaptively adjusts user opinions based on the group's opinion distribution, guiding them toward a target opinion with minimal cost. Borkar et al. ~\cite{borkar2021opinion} 
applied RL to identify the most influential users through agent interactions, addressing the challenge of unknown network topology. 
These methods fully utilize RL's capability to navigate complex environments, identifying optimal strategies through agent-environment interactions. However, while these methods primarily focus on cost minimization, our approach emphasizes optimizing the efficiency of consensus.

In recent years, several studies have examined the role of interaction intensity between users in the evolution of social networks. Nugent et al.~\cite{nugent2024steering} proposed a strategy for opinion guidance by regulating the strength of user interactions. 
Bolzern et al.~\cite{bolzern2020opinion} examined how guiding the content users are exposed to affects opinion formation.
While these studies show that opinions can be influenced by altering interaction behaviors, they do not fully account for the hierarchical relationships among users or the efficiency of opinion guidance mechanisms.

\section{Preliminaries}

\subsection{Opinion Dyanmics Environment}
In the initial opinion dynamics environment, we define $E=\{e_1, e_2, \\ \cdots, e_n\}$ as the set of $n$ experts, and $V=\{v_1, v_2, \cdots, v_m\}$ as the set of $m$ non-experts. Let $x_{e_i}(k)$ and $x_{v_i}(k)$ denote the opinion of expert $e_i$ and non-expert $v_i$ at time $k$, respectively. For clarity, we assume that the initial opinions of the non-experts satisfy $x_{v_1}(0) \leq x_{v_2}(0) \leq \cdots \leq x_{v_m}(0)$, so do the experts. In the context of this study, experts are defined as individuals with an authoritative position, a status derived from their social attributes and public recognition, making their position relatively fixed and irreplaceable in the short term. Therefore, this paper focuses on opinion changes within a specific topic over a short period, during which the identity and influence of experts are considered stable.




\subsection{MARL under Opinion Dynamics Scenarios}
\label{sec:pre-marl}
Markov Decision Processes (MDPs) are the basis of modeling and analysis in MARL.
According to the analysis in the previous section, we extend the definition of MDPs in MARL proposed by Zhang \textit{et al.} \cite{DBLP:conf/icml/ZhangYL0B18}, and defined a new MDP for the opinion dynamics under social networks in Eq. \eqref{eq:mdmp}. 
The brief introduction is given as follows:
\begin{equation}
	\label{eq:mdmp}
	\left(
		\left\{S_{v_i}\right\},
		\left\{A_{v_i}\right\}, 
		P,
		\left\{R_{v_i}\right\},
		\left\{L(k)\right\}_{k\geqslant 0} \in \mathbb{R}^{N \times N}
	\right),
\end{equation}
where 
$\left\{S_{v_i}\right\}$ represents the local state space perceived by agent $v_i$, 
$\left\{A_{v_i}\right\}$ represents the local action space made by agent $v_i$, 
$\left\{L(k)\right\}_{k\geqslant 0}$ indicates the communication topology matrix of the system at time $k$. If there exists a communication channel from agent $v_i$ to agent $v_j$, then $l_{ij}$ is $1$,otherwise, $l_{ij}$ is $0$.
Then, the local reward function obtained by each agent $i$ is $R_{v_i}: S \times A \rightarrow R$, the state transition probability of MDP is $P: S \times A \times S \rightarrow [0,1]$.

The Markov decision process defined in this paper can be expressed as follows: 
agent $v_i$ perceives the local observation state $S_{v_i}(k)$ at time $k$. 
Then, $v_i$ get the action probability distribution $A_{v_i}(k)$ according to the shared policy parameters $\pi_{\theta}$. Based on the $A_{v_i}(k)$, we update the communication matrix $L(k)$.
Meanwhile, $v_i$ receives its local instant rewards $r_{v_i}(k)$ based on the reward function.

We represent the state, action, and reward generated by each agent during its interaction with the environment in a training round as $\tau_{v_i}= \left(S_{v_i}(0),A_{v_i}(0),r_{v_i}(0),\cdots,S_{v_i}(H), A_{v_i}(H),r_{v_i}(H)\right)$, where $H$ denotes the length of the trajectory sequence. 
In the process of interaction with the environment, agents constantly learn and adjust their strategy function $\pi_{\theta}(A_{v_i}(k) | S_{v_i}(k))$ to maximize the expected cumulative $\overline{R}_{\theta}$ return in the future:
\begin{equation}
	\label{eq:cum_return}
	\overline{R}_{\theta} = \sum_{\tau_{v_i}}{P\left( \tau _{v_i} | \theta \right) R\left( \tau _{v_i} \right)},
\end{equation}
where $R\left( \tau_{v_i} \right)$ is the total reward obtained from $\tau_{v_i}$, and $P\left( \tau _{v_i} | \theta \right)$ is the probability that the trajectory occurs. 

\subsection{Problem Formulation}
Consider a social network system comprising $n$ experts and $m$ non-epxerts. Each user's initial opinion $x_i(0)$ is distributed within the interval $\left[x_{min},x_{max}\right]$. 
The system aims to guide all users' opinions to converge to a global consensus goal $U$ within $T$ finite time steps through non-intrusive interventions, such that for any user:
\begin{equation}
\lim_{t \to T} x_i(t) = U 
\quad \text{s.t.} \quad 
\max_{j} \lvert x_i(t) - x_j(t) \rvert \leq \omega
\label{eq:conditional_converge},
\end{equation}
where $\omega$ denotes consistency threshold. 

\section{Methodology}
\subsection{Overall Framework}
\begin{figure}[ht]
\centering
\includegraphics[width=\linewidth]{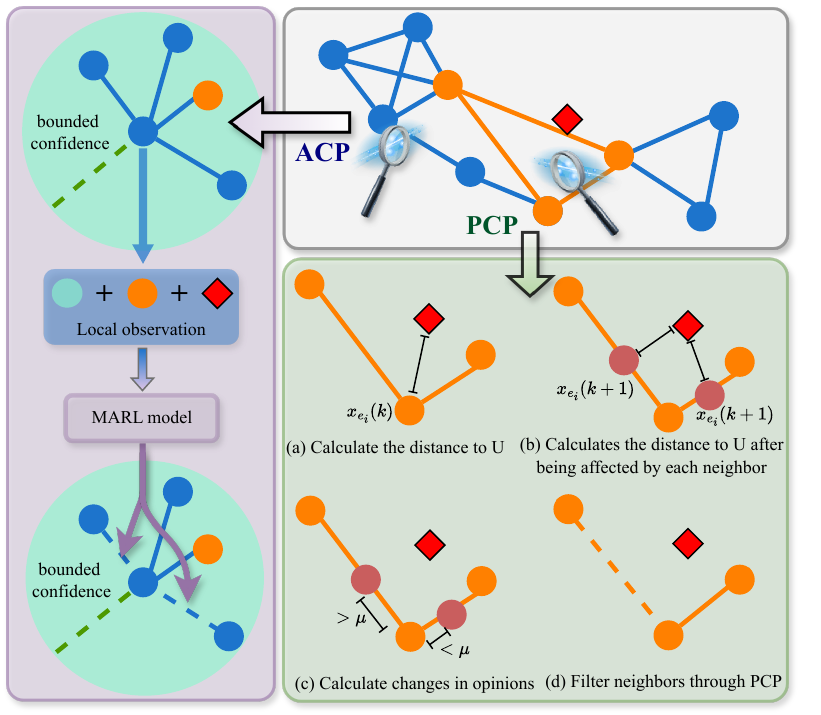}
\caption{
    \ours\ framework. At the start of each training round, a global goal is set and users are categorized into non-experts and experts. PCP filters neighbors for experts based on the minimum disagreement, while ACP filters neighbors for non-experts based on MARL for long-term planning.
}
    \Description{..}
    \label{fig:framework}
\end{figure}

To address the issue of guiding users' opinions towards achieving an inclusive consensus on the goal through neighbor filtering, we propose \ours\ for multi-agent opinion guidance (\figurename~\ref{fig:framework}). It consists of three phases: environment initialization, user communication, and policy exploitation and update.

\textbf{Environment initialization phase.} 
In each training round, we set a global consensus goal and categorized the users into two roles: experts and non-experts. Each role follows a distinct opinion dynamics model that accounts for different levels of stubbornness. 

\textbf{User communication phase.} 
Users engage in role-specific interaction patterns: PCP assigns experts' neighbors based on minimal opinion divergence, propagating global goal influence to non-experts through expert consensus. Meanwhile, ACP dynamically prunes communication channels via MARL-driven balancing of local stability and global consensus, optimizing non-expert interactions through long-term planning strategies.
The communication process continues until it either reaches the maximum time step $T$ or achieves the consensus threshold $\omega$ (see Sections ~\ref{sec:ou_exp} and ~\ref{sec:ou_np}).

\textbf{Policy exploitation and updating phase.} First, the reward obtained by each non-expert in the round is calculated based on the reward function. 
Then, the adaptive attention network within the ACP is trained and optimized using a policy gradient algorithm, learning a policy function that maximizes the expected reward (see Section ~\ref{sec:rw_pg}). 

\subsection{Opinion Updating for Experts}
\label{sec:ou_exp}
Expert users on social media exhibit stable opinions and less susceptible to opinion shifts than non-experts~\cite{zhao2018understanding,casalo2020influencers}. While experts influence non-experts through opinion leadership, their communication largely remains within expert networks. 
This creates a hierarchical information structure, with experts acting as discourse anchors, guiding opinions while remaining isolated from non-expert perspectives.

To formalize these behavioral patterns, we propose PCP coupled with a stubbornness dynamics model. In particular, PCP prefers to establish communication channels among experts with similar perspectives, strategically amplifying opinion exchange within ideologically proximate clusters to accelerate alignment processes while preserving experts' core epistemic positions.


At moment $k$, we find the neighbors $G_{e_{i}}(k)$ that effectively contribute to the $e_i$ to move closer to the global goal by using Eq. \eqref{eq:nei_exp_i}:
\begin{equation}
    \label{eq:nei_exp_i}
    G_{e_{i}}(k) = \left\{e_{j} : e_j \in E , \lvert\frac{x_{e_i}(k)+ x_{e_j}(k)}{2}- U\rvert \leq \lvert x_{e_i}(k) - U \rvert \right\}.
\end{equation}

Following this, if expert $e_i$ has no neighbors that can help it to converge towards the goal, its opinion remains unchanged during this time step, i.e., $x_{e_i}(k+1) = x_{e_i}(k)$. Otherwise, $e_i$ updates its opinion using the stubbornness dynamics model for experts, as described in Eq. \eqref{eq:nei_exp_i+1}:
\begin{equation}
    \label{eq:nei_exp_i+1}
    x_{e_i}(k+1) =\mathrm{\phi}_{e_{i}} \cdot x_{e_i}(k) + \left(1-\phi_{e_{i}}\right) \cdot \frac{\sum_{j \in{\hat{G}}_{e_i}(k)}x_{e_j}(k)}{\left|\hat{G}_{e_i}(k)\right|},
\end{equation}
where $\mathrm{\phi}_{e_{i}}$ represents the stubbornness level of expert $e_{i}$ and $\hat{G}_{e_i}(k)$ is a filtered set of neighbors that have relatively small differences of opinion with $e_i$ and are more likely to be received. The steps to obtain $\hat{G}_{e_i}(k)$ are as follows:





\begin{description}
\item[Step $1$] Select the nearest neighbour $e_{j}$ from $G_{e_{i}}(k)$ and place it in  $\hat{G}_{e_i}(k)$.
\item[Step $2$] Calculation of $x_{e_i}(k+1)$ according to Eq. \ref{eq:nei_exp_i+1}.
\item[Step $3$] If $\left|x_{e_i}(k+1) - x_{e_i}(k)\right| \geq \mu$, the selection of $\hat{G}_{e_i}(k)$ is complete; otherwise, return to Step $1$.
\end{description}


The parameter $\mu$ represents the threshold for opinion change among experts, controlling the extent of shifts in their opinions. This control is necessary because the number of experts is much smaller compared to non-experts, and within the same opinion range, experts often display more disagreement. When experts communicate, their opinions can shift more significantly, despite their generally higher stubbornness. If experts' opinions change too quickly, non-experts may struggle to keep up, potentially losing the guidance provided by the experts.

To manage the influence range of experts, we divide the users in the environment into subgroups. Initially, each non-experts selects the expert whose opinion is closest to their own as their guide. Non-experts who follow the same expert as well as this expert together form a subgroup, and non-experts are only influenced by the experts within that subgroup, unaffected by experts from other groups. We use a set $LE_{v_i}(k)$ to represent the experts that non-expert $v_i$ follows at time $k$.

As users communicate, their opinions evolve over time, leading to changes in subgroup composition.
Subgroups merge when the difference between the average opinions of the experts in any two subgroups is less than or equal to the merge threshold $\beta$.
When this condition is met, the experts and non-experts from both subgroups combine into a new group. Gradually, as differences in expert opinions are reconciled, the environment gradually consolidates into a single subgroup. In this final subgroup, all non-experts are influenced by all experts.

\subsection{Opinion Updating for Non-experts} 
\label{sec:ou_np}
Compared to experts with stable opinions, non-expert users in social networks exhibit greater opinion flexibility, making their opinions more susceptible to interactions. Additionally, in dynamic social topologies, non-experts typically interact with a larger set of neighbors.
To enhance consensus efficiency, we propose a coupled framework integrating ACP with a dynamic non-expert stubbornness model. Specifically, ACP adaptively filters interaction partners based on neighborhood distribution and consensus goals, balancing local group cohesion and global goal orientation.
During opinion evolution, non-expert' neighborhoods dynamically shift with system states, creating uncertainty in interaction partners' quantity and attributes. This demands decision models to handle real-time evolving local goals (linked to expert opinions) and variable-length neighborhood inputs. 
To address these challenges, ACP employs a bidirectional LSTM-based decision model that not only effectively captures the dynamic influence of neighbors but also adaptively processes neighborhood inputs of varying scales through its bidirectional encoding mechanism.

The model inputs must consider the agents' current consensus goal and the distribution of their neighbors.
The set $N_{v_{i}}(k) = \{v_j : v_j \in V, \lvert x_{v_i}(k)-x_{v_j}(k)\rvert< r_c\}$ defines the neighbors of agent $v_i$ at time $k$, where $r_c$ represents the bounded confidence level. We determine the local goal $O_{v_{i}}(k)$ through:
\begin{equation}
    \label{eq:get_TA}
    O_{v_{i}}(k) = \frac{\sum_{e_j \in LE_{v_i}(k)}x_{e_{j}}(k)}{\left|LE_{v_i}(k)\right|}.
\end{equation}

Once the local consensus goal of $v_i$ is determined, we first calculate the distance $d^O_{v_i}(k)$ from $v_i$
to the local goal and the distance $d^U_{v_i}(k)$ to the global goal, as described in the following equation:
\begin{equation}
    \label{eq:d_local}
    d^O_{v_i}(k) =  \left| x_{v_i}(k) - O_{v_i}(k) \right|,\
    d^U_{v_i}(k) =  \left| x_{v_i}(k) - U \right|.
\end{equation}

Next, we process the neighbor information perceived by $v_i$. $\bm{o}$ represents the distribution of neighbors $N_{v_{i}}(k)$ relative to the local goal, while $\bm{u}$ represents the distribution relative to the global goal, as shown below:
\begin{equation}
    \label{eq:distr_gl}
    \bm{o} = \text{Shuffle}\left( \left\{\frac{d^O_{v_j}(k)}{d^O_{v_i}(k)}\right\}_{v_{j} \in N_{v_i}} \right),\
    \bm{u} = \text{Shuffle}\left( \left\{\frac{d^U_{v_j}(k)}{d^U_{v_i}(k)}\right\}_{v_{j} \in N_{v_i}} \right).
\end{equation}

In Eq.~\eqref{eq:distr_gl}, if the element in $\bm{o}$ or $\bm{u}$ is less than $1$, it indicates that agent $v_{j}$ contributes to the consensus of $v_{i}$ toward the local or global goal. Conversely, if the element in $\bm{o}$ or $\bm{u}$ is 1 or greater, it means $v_{j}$ does not contribute to the consensus of $v_{i}$ toward these goals. To reduce the bias caused by the indexing of agents, we apply a shuffle function to randomly reorder the elements in $\bm{o}$ and $\bm{u}$.
It is worth noting that if $d^U_{v_i}(k)$ or $d^O_{v_i}(k)$ is less than or equal to $0.1$, we make then equal to $0.1$ to control the range of values in Eq.~\eqref{eq:distr_gl}.

Finally, we aggregate this information to obtain $S_{v_i}(k)$ as an input to the LSTM model. The specific representation of $S_{v_i}(k)$ is described as follows:
\begin{equation}
    \label{eq:sik}
    S_{v_i}(k) =\left[\bm{o},\bm{u},d^O_{v_i}(k),d^U_{v_i}(k)\right].
\end{equation}

After passing through the LSTM network, the output $A_{v_i}(k)\sim\pi_{\theta_{i}}^j $ is the action probability of the selection each neighbor, $\pi_{\theta_{i}}^j$ represents the probability of agent $i$ choosing agent $j$ under policy $\pi_{\theta}$. 
Therefore, the output layer of the LSTM network contains the SoftMax activation function, which ensures that the combined probability of all actions is $1$.

To help the model to be fully explored, we adopt the strategy of $\varepsilon-greedy$. During the communication process, 
agents occasionally take opposite actions to increase exploration in the environment.
$\varepsilon$ will vary periodically with the increase in the number of rounds.

Based on the model output, we update the communication topology matrix $L(K) = \left[l_{ij}(k)\right]$ by aggregating the neighbors and non-neighbors of the agents in index order.
Then, we set the average value $\hat{\pi}_{\theta_{i}}$ of $a_{v_{i}}(k)\sim\pi_{\theta_{i}}^j$ as the standard to assign to neighbors.
\begin{equation}
    \label{eq:topo_new}
    l_{ij}= \left\{
    \begin{aligned}
        1,&\ v_j \in N_{v_i}(k)\ and\ \pi_{\theta_{i}}^j \geq \hat{\pi}_{\theta_{i}}\\
        0,&\ v_j \notin N_{v_i}(k)\ or\ \pi_{\theta_{i}}^j < \hat{\pi}_{\theta_{i}}\\
    \end{aligned}
    \right. 
\end{equation}

At this point, the agent's neighbors shift from those within the bounded confidence range to those filtered by ACP. We update the opinions of non-experts based on $L(k)$. During this process, each agent is influenced by its neighbors and also by the local goal, i.e., the experts it follows. The influence from experts and neighbors is controlled by parameters $p$ and $q$, as shown in Eq.~\eqref{eq:C_vi}:

\begin{equation}
    \label{eq:C_vi}
    C_{v_i}(k) = p \cdot O_{v_{i}}(k) + q \cdot \frac{\sum_{v_j \in V} l_{ij} \cdot x_{v_j}(k)}{\sum_{v_j \in V} l_{ij}}.
\end{equation}

Then, the stubbornness dynamic model for non-experts is proposed in Eq. \eqref{eq:normal_people}:
\begin{equation}
    \label{eq:normal_people}
     x_{v_i}(k+1) = \left\{
     \begin{aligned}
     \mathrm{\phi}_{v_{i}} \cdot x_{v_i}(k) + (1-\mathrm{\phi}_{v_{i}})  \cdot C_{v_i}(k) ,&\ if\ N_{v_i}(k) \neq \varnothing 
     \\
     x_{v_i}(k) ,&\ if\ N_{v_i}(k) = \varnothing\\
     \end{aligned}
     \right. ,
\end{equation}
where $\mathrm{\phi}_{v_{i}}$ represents the stubbornness level of non-experts $v_i$, $C_{v_i}(k)$ represents the opinion received by the agent $v_i$ at time $k$. 

\subsection{Reward Design and MARL Optimization}
\label{sec:rw_pg}
To enable non-expert users to balance local and global consensus goals during long-term opinion evolution, we design a dual-reward mechanism that integrates two complementary goals, as shown in Eq.~\eqref{eq:reward_k}. This mechanism aims to guide non-experts toward global objective convergence while maintaining local group connectivity, thereby preventing premature fragmentation caused by limited confidence. Given that the immediate impacts of communication channels may manifest only after multiple steps, we employ policy gradient optimization to associate current actions with long-term benefits by maximizing cumulative rewards, effectively mitigating negative effects from myopic decision-making.
\begin{equation}
    \label{eq:reward_k}
     r_{v_i}(k) = g_1(\cdot) + g_2(\cdot)
\end{equation}


\textbf{Local Consensus Reward}: $g_1(\cdot)$ guides agents toward local consensus goal by ensuring users maintain communication ranges with their subgroups, thereby preventing edge users from isolation:

\begin{equation}
    \label{eq:g1_1}
     g_1 = \left\{
     \begin{aligned}
     \frac{d^O_{v_i}(k) - d^O_{v_i}(k+1)}{d^O_{v_i}(k)}, if d^O_{v_i}(k) \geq \xi_l
     \\
     \frac{\xi_l - d^O_{v_i}(k+1)}{\xi_l}, if d^O_{v_i}(k) < \xi_l \\
     \end{aligned}
     \right. ,
\end{equation}
where $\xi_l$ denotes the stabilization threshold from the agent to the global goal. 

When the distance from the agent to the local goal is greater than or equal to $\xi_l$, the local consensus reward is triggered. At this point, $g_1$ indicates the magnitude of the change in the opinion of the agent at moment $k+1$ relative to the moment $k$. 
If $g_1 \geq 0$, it indicates that the agent is approaching the local goal through the communication of moment $k$, and the reward value is proportional to the proximity distance. If $g_1 < 0$, it means that the agent is moving away from the local local through communication at moment $k$. At this point, $g_1$ becomes a penalty, and the penalty value is proportional to the distance away.

The local stabilization reward is activated when the agent's distance to the local goal is less than $\xi_l$. At this point, since the agent's opinion is sufficiently close to the local goal, we aim for the agent to continue fluctuating around it. 
If $d^O_{v_i}(k+1)$ does not exceed $\xi_l$, can be rewarded. Otherwise, if $d^O_{v_i}(k+1)$ exceeds $\xi_l$, it becomes penalized.

\textbf{Global Consensus Reward}: $g_2(\cdot)$ drives agents to converge on global consensus goal, prioritizing long-term consensus guarantees over short-term cohesion, thus preventing subgroup fragmentation:
\begin{equation}
    \label{eq:g2}
     g_2 = \left\{
     \begin{aligned}
     \frac{d^U_{v_i}(k) - d^U_{v_i}(k+1)}{d^U_{v_i}(k)}, if d^U_{v_i}(k) \geq \xi_g
     \\
     \frac{\xi_g - d^U_{v_i}(k+1)}{\xi_g}, if U^O_{v_i}(k) < \xi_g \\
     \end{aligned}
     \right. ,
\end{equation}
where $\xi_g$ denotes the stabilization threshold from the agent to the global goal. The meaning of $g_2$ is basically the same as $g_1$, but the goal of $g_2$ changes from a local goal to a global goal.

Our objective is to maximize the expectation of $\overline{R}_{\theta}$.
We sample the trajectories of all agents in a round and estimate the gradient by averaging the approximate values:
\begin{equation}
	\label{eq:gradient4}
	\nabla _{\theta}\overline{R}_\theta \approx \frac{1}{m}\sum_{i=1}^m{\sum_{k=0}^H{\nabla _{\theta}\log \pi _{\theta}\left( A_{v_i}\left( k \right) |S_{v_i}\left( k \right) \right)} \cdot R\left( \tau _{v_i} \right)}.
\end{equation}

Then, we use the gradient descent method with a learning rate $\alpha$ to update the policy parameters $\theta$:
\begin{equation}
	\label{eq:update}
	\theta \gets \theta +\alpha \nabla _{\theta}\overline{R}_{\theta}.
\end{equation}

\section{Experiment}
\subsection{Experimental Settings}
In this section, extensive experiments were conducted to assess the effectiveness of \ours\ in guiding consensus toward a target opinion. However, acquiring real-world dynamic process data for closed-loop public opinion guidance remains challenging, as most publicly available graph datasets contain static opinion data, which fails to capture user evolution after intervention. To address this, many studies~\cite{chen2021Influence,cicolani2024opinion,nugent2023evolving} rely on synthetic datasets, which allow for better control of experimental variables and deeper analysis of model behavior. This study similarly uses synthetic datasets to ensure experimental reproducibility and the generalizability of findings. Experiments were conducted across various benchmarks, covering different opinion ranges and agent counts.

                
In order to validate the effectiveness and superiority of our \ours, we evaluate the performance using the following metrics:
\begin{itemize}
\item Consensus Cluster (CC): 
Number of opinion clusters formed at consensus threshold or maximum steps.
Lower CC implies greater consensus, indicating poorer divergence.

\item Consensus Step (CS): 
Steps required for the system to reach the consensus threshold. 
Lower CS indicates less iterations needed for users to achieve consensus, reflecting higher guidance efficiency.

\item System Consensus Deviation (SCD): 
Average deviation of agent opinions from the global goal at termination. 
Lower SCD indicates a smaller deviation of the final consensus result from the goal, demonstrating higher guidance accuracy.
\end{itemize}




\subsection{Comparison methods}
We compare the opinion guidance performance of \ours\ with the following baseline methods: 
\begin{itemize}
\item Common-Neighbor (CNR)~\cite{WANG2015180}: Randomly selects users outside the current user's perception range and adds them to their neighbors.

\item Group-Pressure (GP)~\cite{CHENG2019121900}: Sets group pressure, causing the opinions of certain users to be directly influenced by external group pressures.

\item Position Weight Allocation (PWA)~\cite{mcquade2019social}: Assigns weights to each user's neighbors based on the average opinion of all users.
\end{itemize}




While CNR and GP are direct intervention methods, PWA does not directly alter user communication but instead allocates weights based solely on the current environment without considering long-term planning. Both the pressure level and the percentage of pressured agents in the GP are set to $0.5$. The number of long-range neighbors and the preference in the neighbor selection in CNR are set to $1$ and $0$, respectively.

\subsection{Comparative Experiments of Consensus Guidance without Experts}
\label{sec:wo_exp}

To verify the effectiveness of our method in facilitating consistent consensus, we first conduct experiments in an initial setting where agents' opinions are uniformly distributed. Considering that traditional consistency methods generally tend to converge at the midpoint of the opinion, $U$ is set to be the midpoint of the opinion range. There is an expert in the environment whose opinion remains unchanged, and the role of the expert is only to provide a local goal for the ACP, i.e., $p=0$ and $q=1$. All results present the mean and variance from $10$ optimal results selected after $100$ randomized experiments.

\begin{table}[!t]
	\caption{Statistics of consistency comparison experiment. The best performances is highlighted with a \colorbox{green!25}{green} background, and our results are \textbf{bolded}.}
	\label{tab:com_noexp}
	\centering
		\begin{tabular}{p{0.13\linewidth}<{\centering}p{0.07\linewidth}<{\centering}p{0.15\linewidth}<{\centering}p{0.22\linewidth}<{\centering}p{0.22\linewidth}<{\centering}}
			\toprule
			\makecell[c]{Init. \\range} & $m$ & Method & \makecell[c]{CC $\downarrow$} & \makecell[c]{CS $\downarrow$} \\
			\midrule
			\multirow{4}*{$[0,4]$} & \multirow{4}*{$20$} & PWA & $1$ & $13.2\pm0.36$ \\
			& &  CNR & $1$ & $16.3\pm0.61$ \\
			& &  GP & $1$ & $14.3\pm0.61$ \\
			& &  \textbf{\ours} & \cellcolor{green!25}{\bm{$1$}} & \cellcolor{green!25}{\bm{$13.0\pm0.40$}}\\
			\hline
			\multirow{4}*{$[0,4]$} & \multirow{4}*{$40$} & PWA & $1$ & $13.8\pm0.56$ \\
			& &  CNR & $1$ & $17.7\pm0.41$ \\
			& &  GP & $1$ & $14.5\pm0.65$ \\
			& &  \textbf{\ours} & \cellcolor{green!25}{\bm{$1$}} & \cellcolor{green!25}{\bm{$13.1\pm0.29$}}\\
			\hline
			\multirow{4}*{$[0,8]$} & \multirow{4}*{$40$} & PWA & $3$ & \cellcolor{green!25}{$15.1\pm0.29$} \\
			& &  CNR & $1.5\pm0.25$ & $34.5\pm0.25$ \\
			& &  GP & $3$ & $15.6\pm 0.24$ \\
			& &  \textbf{\ours} & \cellcolor{green!25}{\bm{$1$}} & \bm{$20.5\pm1.25$} \\
			\hline
			\multirow{4}*{$[0,8]$} & \multirow{4}*{$80$} & PWA & $3$ & \cellcolor{green!25}{$16.2\pm0.36$} \\
			& &  CNR & $2.7\pm0.21$ & $35\pm0$ \\
			& &  GP & $3$ & $17.3\pm 0.61$ \\
			& &  \textbf{\ours} & \cellcolor{green!25}{\bm{$1$}} & \bm{$22.0\pm2.80$} \\
			\bottomrule
		\end{tabular}
\end{table}
\begin{figure}[!t]
\centering
\subfloat{
    \includegraphics[width=0.6\linewidth]{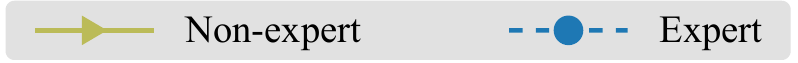}
}
\setcounter{subfigure}{0}
\subfloat[PWA: 3 clusters]{
    \includegraphics[width=0.48\linewidth]{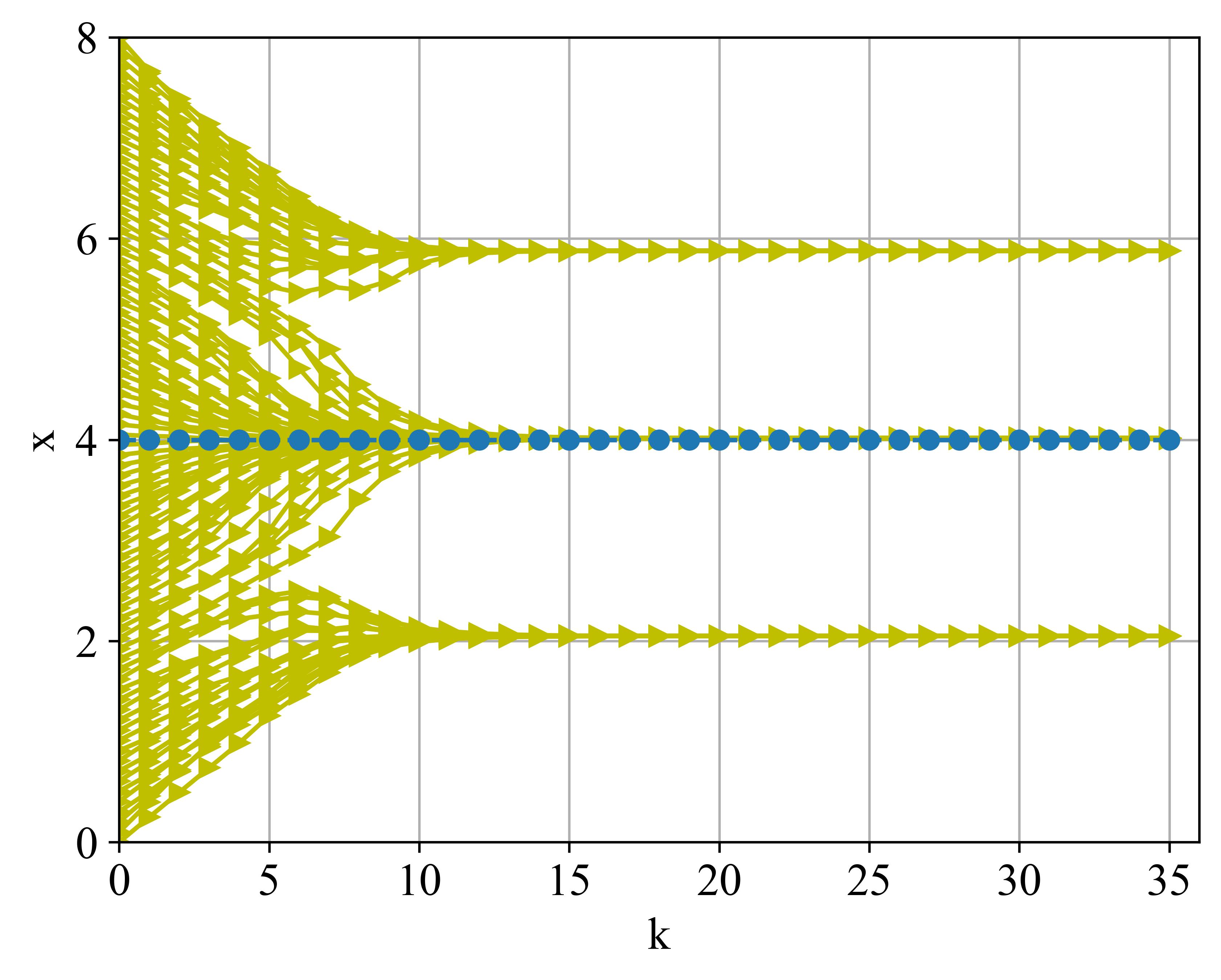}
}
\subfloat[CNR: 2 clusters]{
    \includegraphics[width=0.48\linewidth]{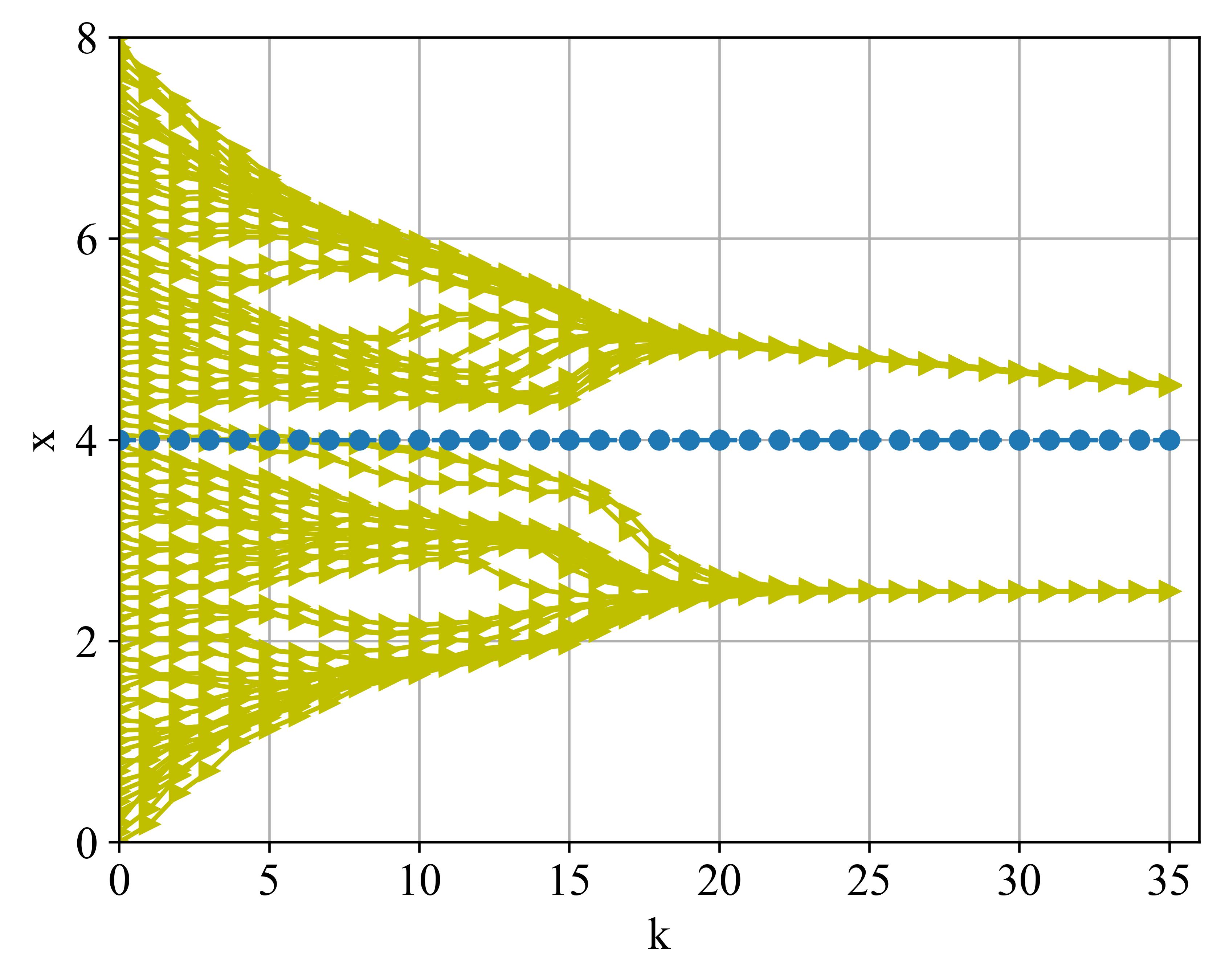}
}\\
\subfloat[GP: 3 clusters]{
    \includegraphics[width=0.48\linewidth]{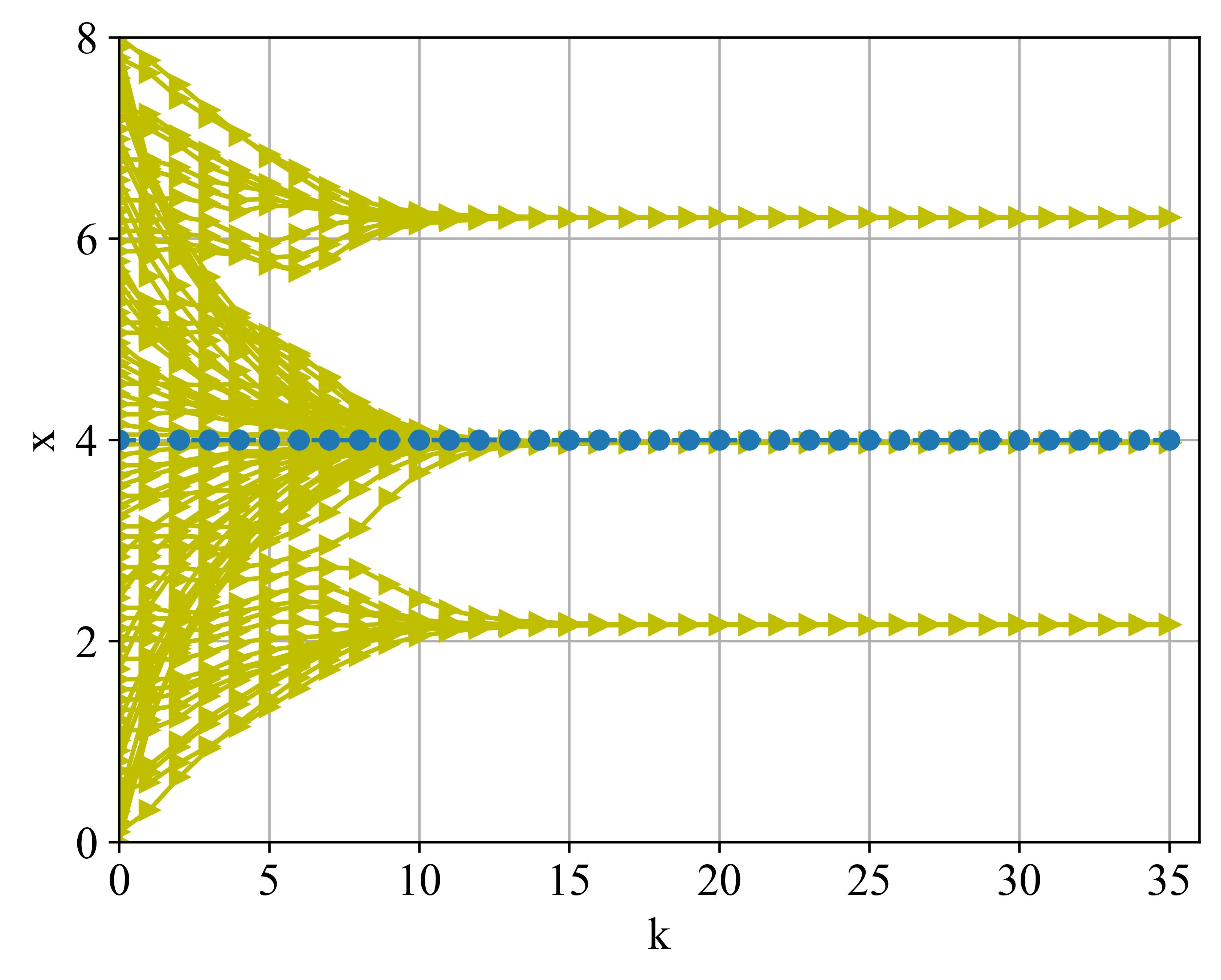}
}
\subfloat[\ours: only 1 cluster]{
    \includegraphics[width=0.48\linewidth]{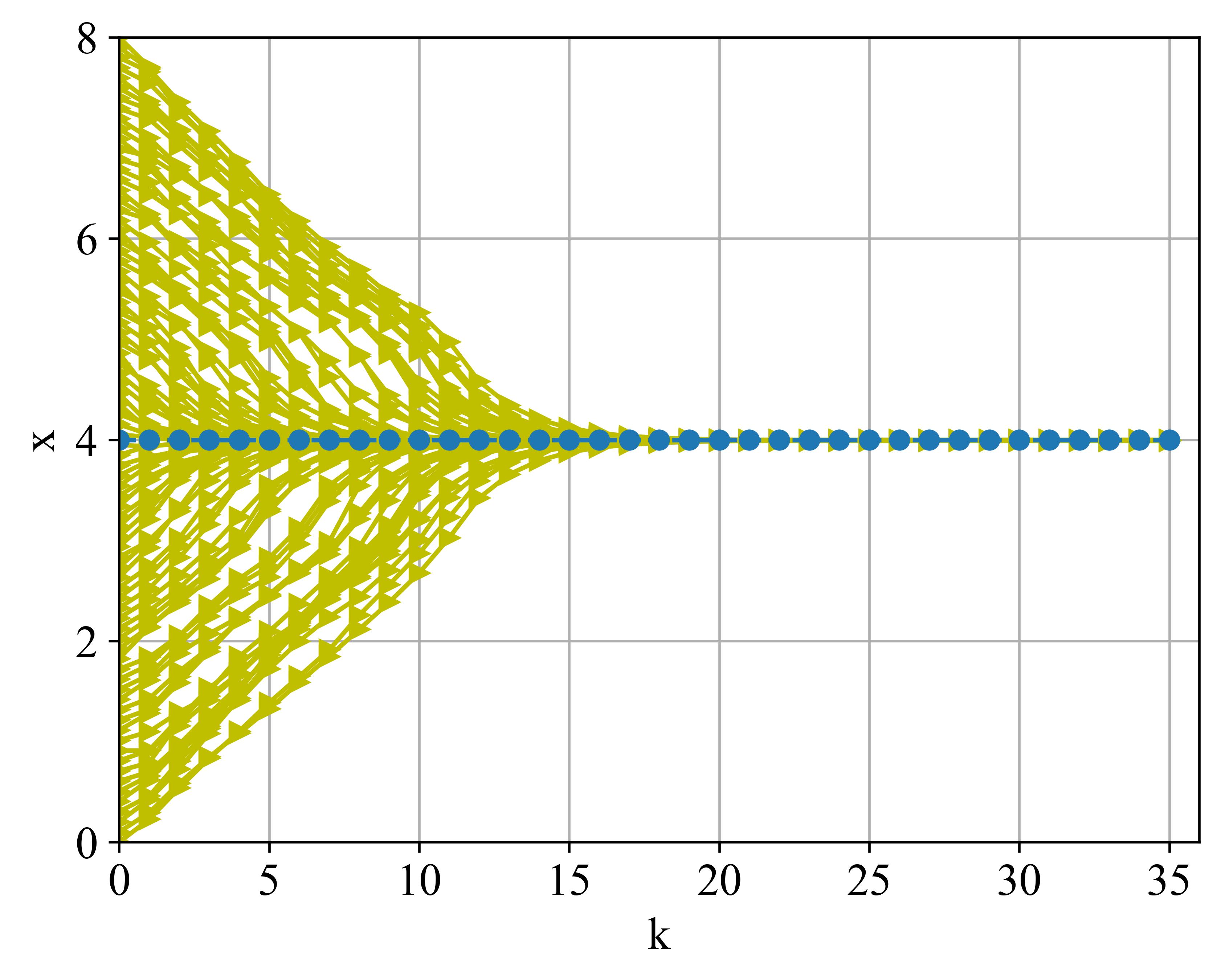}
}
\caption{Illustration of opinion evolution processes without expert influence ($x_{v_i}(0) \in [0,8]$ and $m = 80$).}
\Description{..}
\label{fig:compare-consensus}
\end{figure}
Table~\ref {tab:com_noexp} summarizes the comparative results of promoting consistency under the same conditions. Overall, our method consistently achieves global consensus within a specified number of steps across different environments, varying opinion ranges, and numbers of agents. 
Within the same opinion range, increasing the number of agents does not significantly affect the number of consensus steps or the consensus clusters. 
However, as the opinion range expands, the number of consensus steps increases noticeably, and the consensus quality declines, indicating that greater divergence in agents' opinions leads to higher costs for achieving consensus.

Within the opinion range of $[0,8]$, we observe that PWA and GP require significantly fewer steps to reach consistency but result in a higher number of consensus clusters. In contrast, CNR, though needing more steps and sometimes failing to achieve full consensus within the maximum allowed steps, shows a clear trend toward consensus and is likely to reach global consensus in the future. As shown in \figurename~\ref{fig:compare-consensus}, PWA and GP fall into local consensus prematurely. Similarly, while CNR shows a consensus trend, it also first reaches local consensus, requiring additional time to achieve global consistency. In comparison, \ours, though not the fastest in terms of consensus speed, avoids local consensus and ultimately achieves global consensus by trading off some speed for accuracy.

\subsection{Comparative Experiments of Opinion Guidance with Experts}
\label{sec:w_exp}
\begin{table*}[!t]
	\caption{Statistics of opinion guidance with experts.}
	\label{tab:com_exp}
	\centering
		\begin{tabular}{p{0.10\linewidth}<{\centering}p{0.03\linewidth}<{\centering}p{0.07\linewidth}<{\centering}p{0.05\linewidth}<{\centering}p{0.10\linewidth}<{\centering}p{0.05\linewidth}<{\centering}p{0.15\linewidth}<{\centering}p{0.14\linewidth}<{\centering}}
			\toprule
			\makecell[c]{Init. range} & $m$ & $x_{e_i}(0)$ & $U$ & Method & \makecell[c]{CC $\downarrow$} & \makecell[c]{CS $\downarrow$} & SCD $\downarrow$\\
			\midrule
			\multirow{4}*{$[0,4]$} & \multirow{4}*{$20$} & \multirow{4}*{$[1,3]$} & \multirow{4}*{$1.5$} & PWA & $1$ & $18.5\pm 4.05$ & $0.258 \pm 3.38e^{-4}$ \\
			&& & & CNR & $1$ & $17.0\pm0.60$ & $0.171 \pm 5.16e^{-4}$\\
			&& & & GP & $1$ & $17.3\pm5.41$ & $0.251 \pm 11.38e^{-4}$\\
			&& & & \textbf{\ours} & \cellcolor{green!25}{\bm{$1$}} & \cellcolor{green!25}{\bm{$13.7\pm0.61$}} & \cellcolor{green!25}{\bm{$0.050 \pm 0.29e^{-4}$}}\\
			\hline
			\multirow{4}*{$[0,4]$} & \multirow{4}*{$40$} & \multirow{4}*{$[1,3]$} &\multirow{4}*{$1.5$}  & PWA & $1$ & $19.4\pm0.84$ & $0.239 \pm 1.92e^{-4}$ \\
			&& & &  CNR & $1$ & $19.5\pm 1.05$ & $0.190 \pm 1.11e^{-4}$\\
			&& & & GP & $1$ & $21.0\pm4.60$ & $0.236 \pm 0.54e^{-4}$\\
			&& & & \textbf{\ours} & \cellcolor{green!25}{\bm{$1$}} & \cellcolor{green!25}{\bm{$13.8\pm0.56$}} &  \cellcolor{green!25}{\bm{$0.050 \pm 3.30e^{-4}$}}\\
			\hline
			\multirow{4}*{$[0,8]$} & \multirow{4}*{$40$}& \multirow{4}*{$[1,3,5,7]$} &\multirow{4}*{$3$}  & PWA & $1$ & $34.3\pm 0.81$ & $0.223 \pm 2.28e^{-4}$ \\
			&& & &  CNR & $1$ & $27.8\pm 1.36$ & $0.182 \pm 23.70e^{-4}$\\
			&& & &  GP & $1$ & $32.3\pm 5.21$ & $0.150 \pm 86.85e^{-4}$ \\
			&& & &  \textbf{\ours} & \cellcolor{green!25}{\bm{$1$}} & \cellcolor{green!25}{\bm{$24.1\pm3.49$}} &  \cellcolor{green!25}{\bm{$0.069 \pm 9.06e^{-4}$}}\\
			\hline
			\multirow{4}*{$[0,8]$} & \multirow{4}*{$80$}& \multirow{4}*{$[1,3,5,7]$} &\multirow{4}*{$3$}  & PWA & $1$ & $35\pm0$ & $0.253 \pm 2.64e^{-4}$ \\
			&& & &  CNR & $1$ & $33.5\pm 0.25$ & $0.160\pm4.64e^{-4}$\\
			&& & & GP & $1$ & $35\pm0$ & $0.377 \pm 0.44e^{-4}$ \\
			&& & &  \textbf{\ours} & \cellcolor{green!25}{\bm{$1$}} & \cellcolor{green!25}{\bm{$25.1\pm1.69$}} &  \cellcolor{green!25}{\bm{$0.077 \pm 0.75e^{-4}$}}\\
			\bottomrule
		\end{tabular}
\end{table*}

We compared the effectiveness of opinion guidance across various opinion ranges and agent numbers. Most existing methods rely on a single expert to directly influence the opinions of non-experts, while our framework operates with multiple experts. To ensure a fair comparison, we applied our proposed hierarchical framework to PWA, GP, and CNR, keeping all other parameters constant. In the experiments assessing the impact of experts' opinions, we set $p=0.1$ and $q=0.9$.

By comparing Tables~\ref {tab:com_noexp} and ~\ref{tab:com_exp}, it can be observed that with the introduction of the influence of experts, all methods can converge into a single cluster within the maximum number of steps. Our method requires the least number of consensus steps in different environments and minimizes the SCD metric when the system reaches the consensus threshold, which means that our method is closest to the global consensus goal when it comes to consensus. 

From Table~\ref {tab:com_exp}, it can be found that the number of consensus steps is longer in the range $[0,4]$ so compared to the case of no expert influence. This is due to the fact that the global consensus goal is not the midpoint of the opinion range. Agents will adjust further towards the goal under the influence of experts, thus slowing down the overall consensus. Our method is also affected to some extent, but the change is minimal.

In the opinion range of $[0,8]$, we observe that while the number of consensus steps increases for all methods, they all ultimately achieve global consistency. Interestingly, unlike the scenario without expert influence, CNR requires fewer consensus steps than PWA and GP. This is because, without experts' guidance, PWA and GP tend to fall into local consensus prematurely, requiring more time for the experts to steer the agents towards the global goal. In contrast, CNR exhibits a natural inclination towards global consensus even without expert influence, making it easier for experts to guide the agents. 
Our method, however, prevents premature local consensus by using long-term planning to manage agent communication, resulting in faster and more efficient global consensus.

\subsection{Opinion Evolution Analysis under Unevenly Distributed Environments}
\label{sec:uneven}
We conducted experiments under the condition of uneven distribution of initial agent opinions to demonstrate the generalizability of our method. 
Under the uneven initial environment, we make the agents randomly distributed in the range of $[0,1]$ and $[2,4]$, respectively. 
The initial opinions of the experts are $0.5$ and $3$.
The number of agents in the range $[0,1]$ is $10$ and the number of agents in the range $[2,4]$ is $30$. $U$ is set to $1.5$ since there are no agents in the range $[2,3]$. 
\begin{figure}[ht]
\centering
\subfloat[Evolution process w/o experts]{
	\label{uf_noexp}\includegraphics[width=0.48\linewidth]{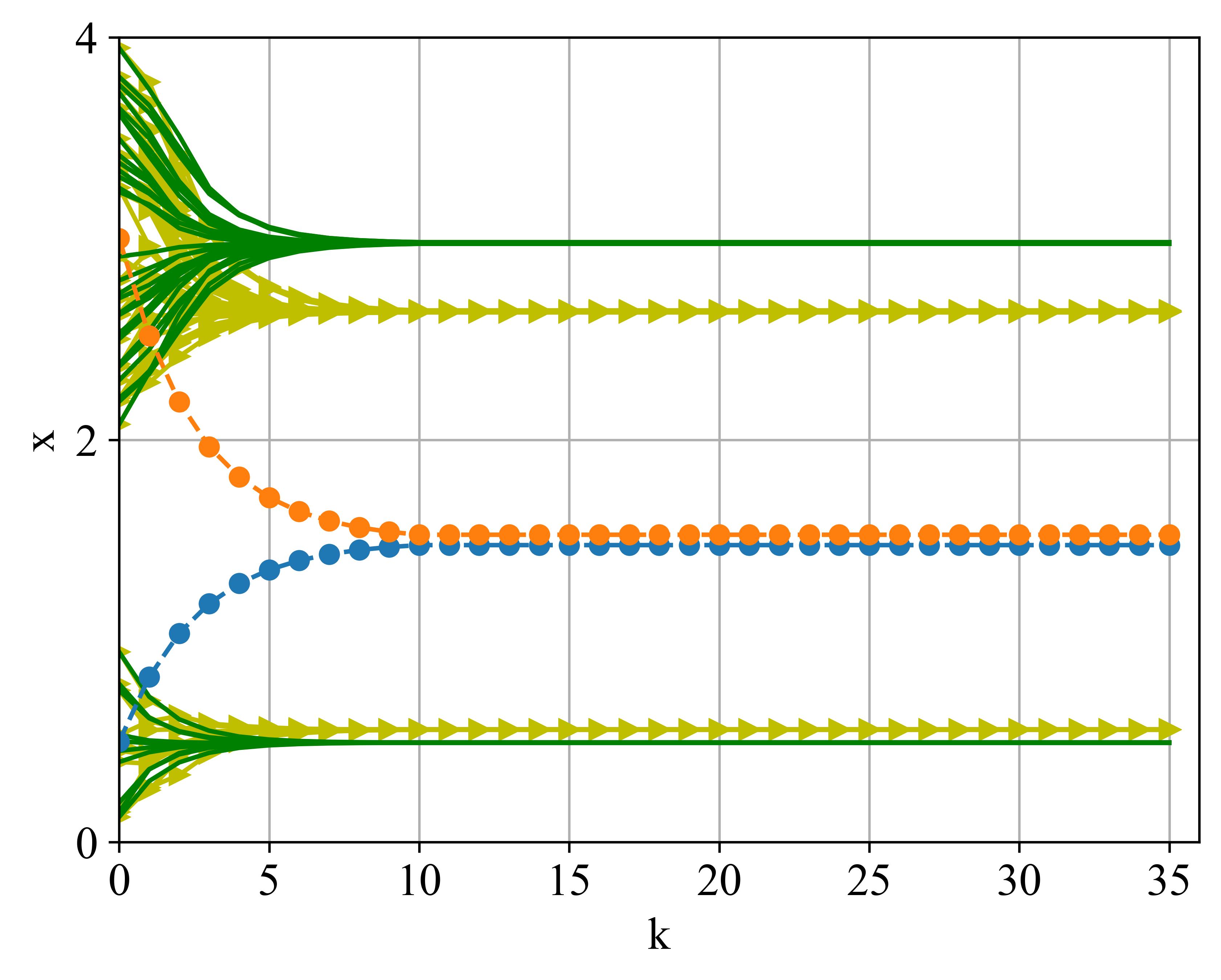}
}
\subfloat[Evolution process w/ experts]{
	\label{uf_exp}\includegraphics[width=0.48\linewidth]{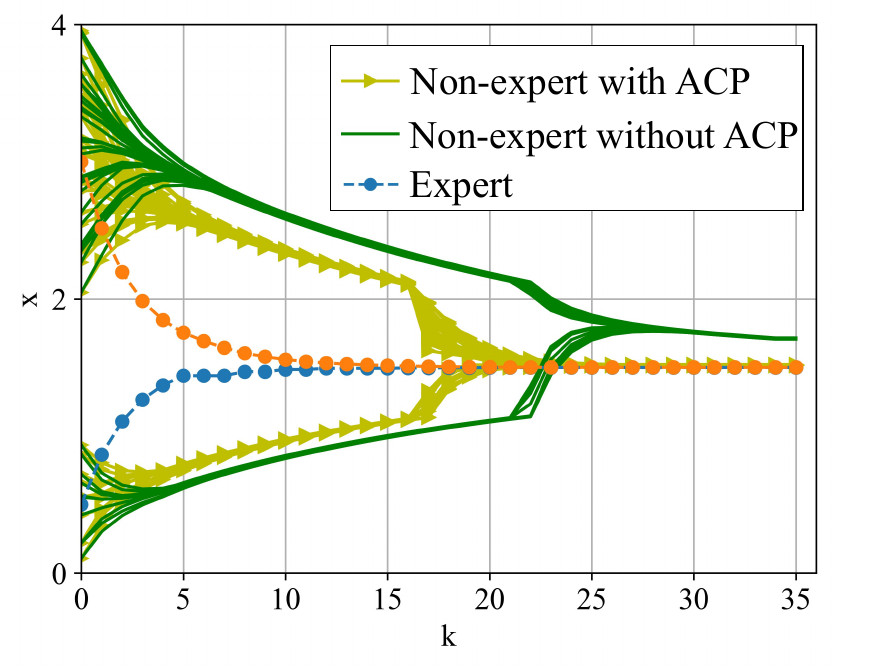}
}
\caption{Opinion evolution processes under uneven opinion distributions.} 
\Description{..}
\label{fig_uf}
\end{figure}


In \figurename~\ref{uf_noexp}, $p=0$ and $ q=1$ (i.e., non-experts will not be directly influenced by experts), we observe that under the action of ACP, the consensus result of the agents is relatively closer to the goal. However, due to bounded confidence, they only achieve local consensus and fail to converge fully to the goal.
In \figurename~\ref{uf_exp}, $p=0.1$ and $q=0.9$ (i.e., non-experts are influenced by experts)  we see that under ACP, non-experts achieve global consensus towards the goal, and the consensus speed is faster compared to scenarios without ACP intervention.
Meanwhile, as shown by the green trajectory, when the opinion distance between two groups is less than $r_c$, the two groups will approach each other and the final result is closer to the one with more agents. However, with ACP intervention, the group with a larger number of agents does not drag the group with a smaller number of agents away, but both of them converge towards the goal. This demonstrates that ACP can effectively combine the global goal and the local information perceived by the agents to make decisions adaptively.

\subsection{Opinion Evolution Analysis under Multi-Expert Conditions}
\label{sec:large_ana}
\begin{figure}[ht]
    \centering
    \includegraphics[width=0.7\linewidth]{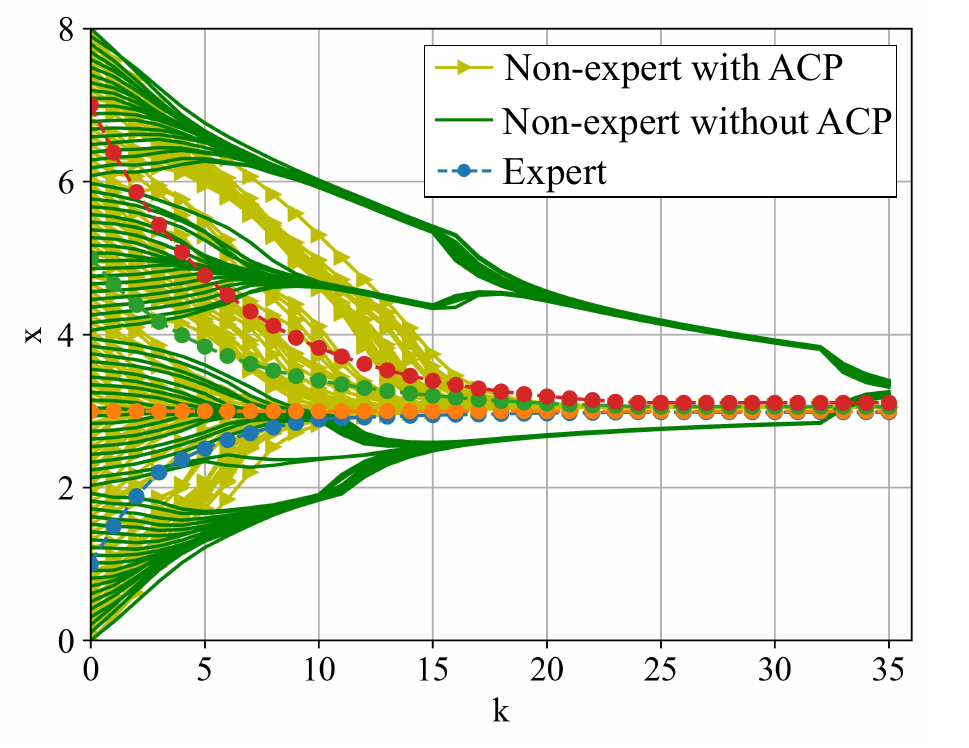}
    \caption{Opinion guidance processes in multi-expert settings.}
    \Description{..}
    \label{fig:big_gen}
\end{figure}

\figurename~\ref{fig:big_gen} illustrates the agents' evolutionary trajectory under the influence of multiple experts within the $[0,8]$ opinion range. Initially, the experts' opinion distribution is $[1,3,5,7]$, with the global consensus goal set to $3$.
It can be found that in the absence of ACP interference, agents are constrained by $r_c$. Non-experts who initially follow the same expert first achieve internal consistency under the expert's influence. Even if some non-experts are initially closer to the global goal than the expert, they may move away from the goal due to the combined influence of the expert and their neighbors. 
After achieving local consensus, agents can only slowly converge to the global goal under expert guidance. This demonstrates that expert influence does not always accelerate consensus to the goal.

Under ACP's influence, both global and local goals are balanced using the local information perceived by the agents. When expert guidance hinders an agent's consensus toward the global goal, ACP compensates by adjusting neighborhood assignments to keep the agent on track. Additionally, considering the effect of bounded confidence, agents that converge too quickly can isolate others. For example, in the lower part of the green trajectory in the $[2,4]$ range, two groups of agents directly reach the goal, but this causes the agents in the $[0,2]$ range to lose their communicative ``bridge'' neighbors, leading to isolated clusters. This issue is avoided with ACP intervention, illustrating the importance of long-term planning. Peripheral agents must move quickly toward the goal, while those closer to the goal should act as ``bridge'' to maintain connectivity and prevent isolation during consensus. 

\section{Conclusion and Discussion}
Social media platforms, while providing engaging content to users, also amplify the polarization of opinions. Traditional methods often directly intervene by introducing external opinions or control functions, which are invasive and not user-friendly. Furthermore, these methods lack a long-term perspective and tend to overly optimize local consensus, leading to divisions at the global level. 
To address these issues, we propose a hierarchical neighbor filtering method (\ours) that considers user roles. \ours\ first introduce a layered communication structure based on role-specific behavioral patterns, categorizing users into experts and non-experts. In the neighbor filtering approach, H-NeiFi proposes a non-intrusive guidance paradigm based on social relationship pruning: PCP and ACP. PCP assigns neighbors to experts based on minimal divergence, while ACP integrates global goal and local information through MARL to dynamically assign neighbors to non-experts.
Experimental results show that our method significantly enhances the system's convergence towards target consensus and effectively promotes broader opinion harmony and the development of cohesive communities.

\bibliographystyle{ACM-Reference-Format}
\bibliography{ref}


\begin{thebibliography}{51}


\ifx \showCODEN    \undefined \def \showCODEN     #1{\unskip}     \fi
\ifx \showISBNx    \undefined \def \showISBNx     #1{\unskip}     \fi
\ifx \showISBNxiii \undefined \def \showISBNxiii  #1{\unskip}     \fi
\ifx \showISSN     \undefined \def \showISSN      #1{\unskip}     \fi
\ifx \showLCCN     \undefined \def \showLCCN      #1{\unskip}     \fi
\ifx \shownote     \undefined \def \shownote      #1{#1}          \fi
\ifx \showarticletitle \undefined \def \showarticletitle #1{#1}   \fi
\ifx \showURL      \undefined \def \showURL       {\relax}        \fi
\providecommand\bibfield[2]{#2}
\providecommand\bibinfo[2]{#2}
\providecommand\natexlab[1]{#1}
\providecommand\showeprint[2][]{arXiv:#2}

\bibitem[Bernardo et~al\mbox{.}(2024)]%
        {bernardo2024bounded}
\bibfield{author}{\bibinfo{person}{Carmela Bernardo}, \bibinfo{person}{Claudio Altafini}, \bibinfo{person}{Anton Proskurnikov}, {and} \bibinfo{person}{Francesco Vasca}.} \bibinfo{year}{2024}\natexlab{}.
\newblock \showarticletitle{Bounded confidence opinion dynamics: A survey}.
\newblock \bibinfo{journal}{\emph{Automatica}}  \bibinfo{volume}{159} (\bibinfo{year}{2024}), \bibinfo{pages}{111302}.
\newblock


\bibitem[Bolzern et~al\mbox{.}(2020)]%
        {bolzern2020opinion}
\bibfield{author}{\bibinfo{person}{Paolo Bolzern}, \bibinfo{person}{Patrizio Colaneri}, {and} \bibinfo{person}{Giuseppe De~Nicolao}.} \bibinfo{year}{2020}\natexlab{}.
\newblock \showarticletitle{Opinion dynamics in social networks: The effect of centralized interaction tuning on emerging behaviors}.
\newblock \bibinfo{journal}{\emph{IEEE transactions on computational social systems}} \bibinfo{volume}{7}, \bibinfo{number}{2} (\bibinfo{year}{2020}), \bibinfo{pages}{362--372}.
\newblock


\bibitem[Borkar and Reiffers-Masson(2021)]%
        {borkar2021opinion}
\bibfield{author}{\bibinfo{person}{Vivek~S Borkar} {and} \bibinfo{person}{Alexandre Reiffers-Masson}.} \bibinfo{year}{2021}\natexlab{}.
\newblock \showarticletitle{Opinion shaping in social networks using reinforcement learning}.
\newblock \bibinfo{journal}{\emph{IEEE Transactions on Control of Network Systems}} \bibinfo{volume}{9}, \bibinfo{number}{3} (\bibinfo{year}{2021}), \bibinfo{pages}{1305--1316}.
\newblock


\bibitem[Borzi and Wongkaew(2015)]%
        {borzi2015modeling}
\bibfield{author}{\bibinfo{person}{Alfio Borzi} {and} \bibinfo{person}{Suttida Wongkaew}.} \bibinfo{year}{2015}\natexlab{}.
\newblock \showarticletitle{Modeling and control through leadership of a refined flocking system}.
\newblock \bibinfo{journal}{\emph{Mathematical Models and Methods in Applied Sciences}} \bibinfo{volume}{25}, \bibinfo{number}{02} (\bibinfo{year}{2015}), \bibinfo{pages}{255--282}.
\newblock


\bibitem[Boxell et~al\mbox{.}(2017)]%
        {boxell2017internet}
\bibfield{author}{\bibinfo{person}{Levi Boxell}, \bibinfo{person}{Matthew Gentzkow}, {and} \bibinfo{person}{Jesse~M Shapiro}.} \bibinfo{year}{2017}\natexlab{}.
\newblock \bibinfo{booktitle}{\emph{Is the internet causing political polarization? Evidence from demographics}}.
\newblock \bibinfo{type}{{T}echnical {R}eport}. \bibinfo{institution}{National Bureau of Economic Research}.
\newblock


\bibitem[Casal{\'o} et~al\mbox{.}(2020)]%
        {casalo2020influencers}
\bibfield{author}{\bibinfo{person}{Luis~V Casal{\'o}}, \bibinfo{person}{Carlos Flavi{\'a}n}, {and} \bibinfo{person}{Sergio Ib{\'a}{\~n}ez-S{\'a}nchez}.} \bibinfo{year}{2020}\natexlab{}.
\newblock \showarticletitle{Influencers on Instagram: Antecedents and consequences of opinion leadership}.
\newblock \bibinfo{journal}{\emph{Journal of business research}}  \bibinfo{volume}{117} (\bibinfo{year}{2020}), \bibinfo{pages}{510--519}.
\newblock


\bibitem[Chen et~al\mbox{.}(2021a)]%
        {chen2021Influence}
\bibfield{author}{\bibinfo{person}{Jia Chen}, \bibinfo{person}{Gang Kou}, \bibinfo{person}{Haomin Wang}, {and} \bibinfo{person}{Yiyi Zhao}.} \bibinfo{year}{2021}\natexlab{a}.
\newblock \showarticletitle{Influence identification of opinion leaders in social networks: an agent-based simulation on competing advertisements}.
\newblock \bibinfo{journal}{\emph{Information Fusion}}  \bibinfo{volume}{76} (\bibinfo{year}{2021}), \bibinfo{pages}{227--242}.
\newblock


\bibitem[Chen et~al\mbox{.}(2024)]%
        {chen2024fully}
\bibfield{author}{\bibinfo{person}{Tongtong Chen}, \bibinfo{person}{Fuyong Wang}, \bibinfo{person}{Meiling Feng}, \bibinfo{person}{Chengyi Xia}, {and} \bibinfo{person}{Zengqiang Chen}.} \bibinfo{year}{2024}\natexlab{}.
\newblock \showarticletitle{Fully distributed consensus of linear multi-agent systems via dynamic event-triggered control}.
\newblock \bibinfo{journal}{\emph{Neurocomputing}}  \bibinfo{volume}{569} (\bibinfo{year}{2024}), \bibinfo{pages}{127129}.
\newblock


\bibitem[Chen et~al\mbox{.}(2021b)]%
        {chen2021opinion}
\bibfield{author}{\bibinfo{person}{Xi Chen}, \bibinfo{person}{Panayiotis Tsaparas}, \bibinfo{person}{Jefrey Lijffijt}, {and} \bibinfo{person}{Tijl De~Bie}.} \bibinfo{year}{2021}\natexlab{b}.
\newblock \showarticletitle{Opinion dynamics with backfire effect and biased assimilation}.
\newblock \bibinfo{journal}{\emph{PloS one}} \bibinfo{volume}{16}, \bibinfo{number}{9} (\bibinfo{year}{2021}), \bibinfo{pages}{e0256922}.
\newblock


\bibitem[Cheng and Yu(2019)]%
        {CHENG2019121900}
\bibfield{author}{\bibinfo{person}{Chun Cheng} {and} \bibinfo{person}{Changbin Yu}.} \bibinfo{year}{2019}\natexlab{}.
\newblock \showarticletitle{Opinion dynamics with bounded confidence and group pressure}.
\newblock \bibinfo{journal}{\emph{Physica A: Statistical Mechanics and its Applications}}  \bibinfo{volume}{532} (\bibinfo{year}{2019}), \bibinfo{pages}{121900}.
\newblock
\showISSN{0378-4371}


\bibitem[Cicolani et~al\mbox{.}(2024)]%
        {cicolani2024opinion}
\bibfield{author}{\bibinfo{person}{Chiara Cicolani}, \bibinfo{person}{Badis Ouahab}, {and} \bibinfo{person}{Cristina Pignotti}.} \bibinfo{year}{2024}\natexlab{}.
\newblock \showarticletitle{Opinion dynamics under common influencer assumption or leadership control}.
\newblock \bibinfo{journal}{\emph{arXiv preprint arXiv:2407.16901}} (\bibinfo{year}{2024}).
\newblock


\bibitem[Cinelli et~al\mbox{.}(2021)]%
        {cinelli2021echo}
\bibfield{author}{\bibinfo{person}{Matteo Cinelli}, \bibinfo{person}{Gianmarco De~Francisci~Morales}, \bibinfo{person}{Alessandro Galeazzi}, \bibinfo{person}{Walter Quattrociocchi}, {and} \bibinfo{person}{Michele Starnini}.} \bibinfo{year}{2021}\natexlab{}.
\newblock \showarticletitle{The echo chamber effect on social media}.
\newblock \bibinfo{journal}{\emph{Proceedings of the National Academy of Sciences}} \bibinfo{volume}{118}, \bibinfo{number}{9} (\bibinfo{year}{2021}), \bibinfo{pages}{e2023301118}.
\newblock


\bibitem[DeGroot(1974)]%
        {degroot1974reaching}
\bibfield{author}{\bibinfo{person}{Morris~H DeGroot}.} \bibinfo{year}{1974}\natexlab{}.
\newblock \showarticletitle{Reaching a consensus}.
\newblock \bibinfo{journal}{\emph{Journal of the American Statistical association}} \bibinfo{volume}{69}, \bibinfo{number}{345} (\bibinfo{year}{1974}), \bibinfo{pages}{118--121}.
\newblock


\bibitem[Dietrich et~al\mbox{.}(2017)]%
        {dietrich2017control}
\bibfield{author}{\bibinfo{person}{Florian Dietrich}, \bibinfo{person}{Samuel Martin}, {and} \bibinfo{person}{Marc Jungers}.} \bibinfo{year}{2017}\natexlab{}.
\newblock \showarticletitle{Control via leadership of opinion dynamics with state and time-dependent interactions}.
\newblock \bibinfo{journal}{\emph{IEEE Trans. Automat. Control}} \bibinfo{volume}{63}, \bibinfo{number}{4} (\bibinfo{year}{2017}), \bibinfo{pages}{1200--1207}.
\newblock


\bibitem[Dittmer(2001)]%
        {dittmer2001consensus}
\bibfield{author}{\bibinfo{person}{Jan~Christian Dittmer}.} \bibinfo{year}{2001}\natexlab{}.
\newblock \showarticletitle{Consensus formation under bounded confidence}.
\newblock \bibinfo{journal}{\emph{Nonlinear Analysis: Theory, Methods \& Applications}} \bibinfo{volume}{47}, \bibinfo{number}{7} (\bibinfo{year}{2001}), \bibinfo{pages}{4615--4621}.
\newblock


\bibitem[Eslami et~al\mbox{.}(2015)]%
        {eslami2015always}
\bibfield{author}{\bibinfo{person}{Motahhare Eslami}, \bibinfo{person}{Aimee Rickman}, \bibinfo{person}{Kristen Vaccaro}, \bibinfo{person}{Amirhossein Aleyasen}, \bibinfo{person}{Andy Vuong}, \bibinfo{person}{Karrie Karahalios}, \bibinfo{person}{Kevin Hamilton}, {and} \bibinfo{person}{Christian Sandvig}.} \bibinfo{year}{2015}\natexlab{}.
\newblock \showarticletitle{" I always assumed that I wasn't really that close to [her]" Reasoning about Invisible Algorithms in News Feeds}. In \bibinfo{booktitle}{\emph{Proceedings of the 33rd annual ACM conference on human factors in computing systems}}. \bibinfo{pages}{153--162}.
\newblock


\bibitem[Friedkin and Johnsen(1990)]%
        {friedkin1990social}
\bibfield{author}{\bibinfo{person}{Noah~E Friedkin} {and} \bibinfo{person}{Eugene~C Johnsen}.} \bibinfo{year}{1990}\natexlab{}.
\newblock \showarticletitle{Social influence and opinions}.
\newblock \bibinfo{journal}{\emph{Journal of mathematical sociology}} \bibinfo{volume}{15}, \bibinfo{number}{3-4} (\bibinfo{year}{1990}), \bibinfo{pages}{193--206}.
\newblock


\bibitem[Gonz{\'a}lez-Bail{\'o}n et~al\mbox{.}(2023)]%
        {gonzalez2023asymmetric}
\bibfield{author}{\bibinfo{person}{Sandra Gonz{\'a}lez-Bail{\'o}n}, \bibinfo{person}{David Lazer}, \bibinfo{person}{Pablo Barber{\'a}}, \bibinfo{person}{Meiqing Zhang}, \bibinfo{person}{Hunt Allcott}, \bibinfo{person}{Taylor Brown}, \bibinfo{person}{Adriana Crespo-Tenorio}, \bibinfo{person}{Deen Freelon}, \bibinfo{person}{Matthew Gentzkow}, \bibinfo{person}{Andrew~M Guess}, {et~al\mbox{.}}} \bibinfo{year}{2023}\natexlab{}.
\newblock \showarticletitle{Asymmetric ideological segregation in exposure to political news on Facebook}.
\newblock \bibinfo{journal}{\emph{Science}} \bibinfo{volume}{381}, \bibinfo{number}{6656} (\bibinfo{year}{2023}), \bibinfo{pages}{392--398}.
\newblock


\bibitem[Hegselmann et~al\mbox{.}(2014)]%
        {hegselmann2014optimal}
\bibfield{author}{\bibinfo{person}{Rainer Hegselmann}, \bibinfo{person}{Stefan K{\"o}nig}, \bibinfo{person}{Sascha Kurz}, \bibinfo{person}{Christoph Niemann}, {and} \bibinfo{person}{J{\"o}rg Rambau}.} \bibinfo{year}{2014}\natexlab{}.
\newblock \showarticletitle{Optimal opinion control: The campaign problem}.
\newblock \bibinfo{journal}{\emph{arXiv preprint arXiv:1410.8419}} (\bibinfo{year}{2014}).
\newblock


\bibitem[Herty and Kalise(2018)]%
        {herty2018suboptimal}
\bibfield{author}{\bibinfo{person}{Michael Herty} {and} \bibinfo{person}{Dante Kalise}.} \bibinfo{year}{2018}\natexlab{}.
\newblock \showarticletitle{Suboptimal nonlinear feedback control laws for collective dynamics}. In \bibinfo{booktitle}{\emph{2018 IEEE 14th International Conference on Control and Automation (ICCA)}}. IEEE, \bibinfo{pages}{556--561}.
\newblock


\bibitem[Kann and Feng(2023)]%
        {kann2023repulsive}
\bibfield{author}{\bibinfo{person}{Claudia Kann} {and} \bibinfo{person}{Michelle Feng}.} \bibinfo{year}{2023}\natexlab{}.
\newblock \showarticletitle{Repulsive bounded-confidence model of opinion dynamics in polarized communities}.
\newblock \bibinfo{journal}{\emph{arXiv preprint arXiv:2301.02210}} (\bibinfo{year}{2023}).
\newblock


\bibitem[Kurz(2015)]%
        {kurz2015optimal}
\bibfield{author}{\bibinfo{person}{Sascha Kurz}.} \bibinfo{year}{2015}\natexlab{}.
\newblock \showarticletitle{Optimal control of the freezing time in the Hegselmann--Krause dynamics}.
\newblock \bibinfo{journal}{\emph{Journal of Difference Equations and Applications}} \bibinfo{volume}{21}, \bibinfo{number}{8} (\bibinfo{year}{2015}), \bibinfo{pages}{633--648}.
\newblock


\bibitem[Ledford(2020)]%
        {ledford2020facebook}
\bibfield{author}{\bibinfo{person}{Heidi Ledford}.} \bibinfo{year}{2020}\natexlab{}.
\newblock \showarticletitle{How Facebook, Twitter and other data troves are revolutionizing social science}.
\newblock \bibinfo{journal}{\emph{Nature}} \bibinfo{volume}{582}, \bibinfo{number}{7812} (\bibinfo{year}{2020}), \bibinfo{pages}{328--331}.
\newblock


\bibitem[Li et~al\mbox{.}(2022)]%
        {li2022applications}
\bibfield{author}{\bibinfo{person}{Tianxu Li}, \bibinfo{person}{Kun Zhu}, \bibinfo{person}{Nguyen~Cong Luong}, \bibinfo{person}{Dusit Niyato}, \bibinfo{person}{Qihui Wu}, \bibinfo{person}{Yang Zhang}, {and} \bibinfo{person}{Bing Chen}.} \bibinfo{year}{2022}\natexlab{}.
\newblock \showarticletitle{Applications of multi-agent reinforcement learning in future internet: A comprehensive survey}.
\newblock \bibinfo{journal}{\emph{IEEE Communications Surveys \& Tutorials}} \bibinfo{volume}{24}, \bibinfo{number}{2} (\bibinfo{year}{2022}), \bibinfo{pages}{1240--1279}.
\newblock


\bibitem[Li et~al\mbox{.}(2020)]%
        {li2020consensus}
\bibfield{author}{\bibinfo{person}{Xianwei Li}, \bibinfo{person}{Yang Tang}, {and} \bibinfo{person}{Hamid~Reza Karimi}.} \bibinfo{year}{2020}\natexlab{}.
\newblock \showarticletitle{Consensus of multi-agent systems via fully distributed event-triggered control}.
\newblock \bibinfo{journal}{\emph{Automatica}}  \bibinfo{volume}{116} (\bibinfo{year}{2020}), \bibinfo{pages}{108898}.
\newblock


\bibitem[Liu et~al\mbox{.}(2024)]%
        {liu2024community}
\bibfield{author}{\bibinfo{person}{Yilu Liu}, \bibinfo{person}{Qingfu Zhang}, {and} \bibinfo{person}{Zhenkun Wang}.} \bibinfo{year}{2024}\natexlab{}.
\newblock \showarticletitle{Community opinion maximization in social networks}.
\newblock \bibinfo{journal}{\emph{IEEE Transactions on Evolutionary Computation}} (\bibinfo{year}{2024}).
\newblock


\bibitem[Lu et~al\mbox{.}(2022)]%
        {lu2022uncovering}
\bibfield{author}{\bibinfo{person}{Yunfei Lu}, \bibinfo{person}{Peng Cui}, \bibinfo{person}{Linyun Yu}, \bibinfo{person}{Lei Li}, {and} \bibinfo{person}{Wenwu Zhu}.} \bibinfo{year}{2022}\natexlab{}.
\newblock \showarticletitle{Uncovering the heterogeneous effects of preference diversity on user activeness: A dynamic mixture model}. In \bibinfo{booktitle}{\emph{Proceedings of the 28th ACM SIGKDD Conference on Knowledge Discovery and Data Mining}}. \bibinfo{pages}{3458--3467}.
\newblock


\bibitem[McQuade et~al\mbox{.}(2019)]%
        {mcquade2019social}
\bibfield{author}{\bibinfo{person}{Sean McQuade}, \bibinfo{person}{Benedetto Piccoli}, {and} \bibinfo{person}{Nastassia Pouradier~Duteil}.} \bibinfo{year}{2019}\natexlab{}.
\newblock \showarticletitle{Social dynamics models with time-varying influence}.
\newblock \bibinfo{journal}{\emph{Mathematical Models and Methods in Applied Sciences}} \bibinfo{volume}{29}, \bibinfo{number}{04} (\bibinfo{year}{2019}), \bibinfo{pages}{681--716}.
\newblock


\bibitem[Monti et~al\mbox{.}(2020)]%
        {monti2020learning}
\bibfield{author}{\bibinfo{person}{Corrado Monti}, \bibinfo{person}{Gianmarco De~Francisci~Morales}, {and} \bibinfo{person}{Francesco Bonchi}.} \bibinfo{year}{2020}\natexlab{}.
\newblock \showarticletitle{Learning opinion dynamics from social traces}. In \bibinfo{booktitle}{\emph{Proceedings of the 26th ACM SIGKDD International Conference on Knowledge Discovery \& Data Mining}}. \bibinfo{pages}{764--773}.
\newblock


\bibitem[Musco et~al\mbox{.}(2018)]%
        {musco2018minimizing}
\bibfield{author}{\bibinfo{person}{Cameron Musco}, \bibinfo{person}{Christopher Musco}, {and} \bibinfo{person}{Charalampos~E Tsourakakis}.} \bibinfo{year}{2018}\natexlab{}.
\newblock \showarticletitle{Minimizing polarization and disagreement in social networks}. In \bibinfo{booktitle}{\emph{Proceedings of the 2018 world wide web conference}}. \bibinfo{pages}{369--378}.
\newblock


\bibitem[Nugent et~al\mbox{.}(2023)]%
        {nugent2023evolving}
\bibfield{author}{\bibinfo{person}{Andrew Nugent}, \bibinfo{person}{Susana~N Gomes}, {and} \bibinfo{person}{Marie-Therese Wolfram}.} \bibinfo{year}{2023}\natexlab{}.
\newblock \showarticletitle{On evolving network models and their influence on opinion formation}.
\newblock \bibinfo{journal}{\emph{Physica D: Nonlinear Phenomena}}  \bibinfo{volume}{456} (\bibinfo{year}{2023}), \bibinfo{pages}{133914}.
\newblock


\bibitem[Nugent et~al\mbox{.}(2024)]%
        {nugent2024steering}
\bibfield{author}{\bibinfo{person}{Andrew Nugent}, \bibinfo{person}{Susana~N Gomes}, {and} \bibinfo{person}{Marie-Therese Wolfram}.} \bibinfo{year}{2024}\natexlab{}.
\newblock \showarticletitle{Steering opinion dynamics through control of social networks}.
\newblock \bibinfo{journal}{\emph{arXiv preprint arXiv:2404.09849}} (\bibinfo{year}{2024}).
\newblock


\bibitem[Okawa and Iwata(2022)]%
        {okawa2022predicting}
\bibfield{author}{\bibinfo{person}{Maya Okawa} {and} \bibinfo{person}{Tomoharu Iwata}.} \bibinfo{year}{2022}\natexlab{}.
\newblock \showarticletitle{Predicting opinion dynamics via sociologically-informed neural networks}. In \bibinfo{booktitle}{\emph{Proceedings of the 28th ACM SIGKDD Conference on Knowledge Discovery and Data Mining}}. \bibinfo{pages}{1306--1316}.
\newblock


\bibitem[Okawa et~al\mbox{.}(2021)]%
        {okawa2021dynamic}
\bibfield{author}{\bibinfo{person}{Maya Okawa}, \bibinfo{person}{Tomoharu Iwata}, \bibinfo{person}{Yusuke Tanaka}, \bibinfo{person}{Hiroyuki Toda}, \bibinfo{person}{Takeshi Kurashima}, {and} \bibinfo{person}{Hisashi Kashima}.} \bibinfo{year}{2021}\natexlab{}.
\newblock \showarticletitle{Dynamic hawkes processes for discovering time-evolving communities' states behind diffusion processes}. In \bibinfo{booktitle}{\emph{Proceedings of the 27th ACM SIGKDD Conference on Knowledge Discovery \& Data Mining}}. \bibinfo{pages}{1276--1286}.
\newblock


\bibitem[Pasqualetti et~al\mbox{.}(2014)]%
        {pasqualetti2014controllability}
\bibfield{author}{\bibinfo{person}{Fabio Pasqualetti}, \bibinfo{person}{Sandro Zampieri}, {and} \bibinfo{person}{Francesco Bullo}.} \bibinfo{year}{2014}\natexlab{}.
\newblock \showarticletitle{Controllability metrics, limitations and algorithms for complex networks}.
\newblock \bibinfo{journal}{\emph{IEEE Transactions on Control of Network Systems}} \bibinfo{volume}{1}, \bibinfo{number}{1} (\bibinfo{year}{2014}), \bibinfo{pages}{40--52}.
\newblock


\bibitem[Peng et~al\mbox{.}(2020)]%
        {peng2020novel}
\bibfield{author}{\bibinfo{person}{Zhinan Peng}, \bibinfo{person}{Jiangping Hu}, \bibinfo{person}{Kaibo Shi}, \bibinfo{person}{Rui Luo}, \bibinfo{person}{Rui Huang}, \bibinfo{person}{Bijoy~Kumar Ghosh}, {and} \bibinfo{person}{Jiuke Huang}.} \bibinfo{year}{2020}\natexlab{}.
\newblock \showarticletitle{A novel optimal bipartite consensus control scheme for unknown multi-agent systems via model-free reinforcement learning}.
\newblock \bibinfo{journal}{\emph{Appl. Math. Comput.}}  \bibinfo{volume}{369} (\bibinfo{year}{2020}), \bibinfo{pages}{124821}.
\newblock


\bibitem[Reynolds-Tylus(2019)]%
        {reynolds2019psychological}
\bibfield{author}{\bibinfo{person}{Tobias Reynolds-Tylus}.} \bibinfo{year}{2019}\natexlab{}.
\newblock \showarticletitle{Psychological reactance and persuasive health communication: A review of the literature}.
\newblock \bibinfo{journal}{\emph{Frontiers in Communication}}  \bibinfo{volume}{4} (\bibinfo{year}{2019}), \bibinfo{pages}{56}.
\newblock


\bibitem[Shang(2022)]%
        {shang2022constrained}
\bibfield{author}{\bibinfo{person}{Yilun Shang}.} \bibinfo{year}{2022}\natexlab{}.
\newblock \showarticletitle{Constrained consensus in state-dependent directed multiagent networks}.
\newblock \bibinfo{journal}{\emph{IEEE Transactions on Network Science and Engineering}} \bibinfo{volume}{9}, \bibinfo{number}{6} (\bibinfo{year}{2022}), \bibinfo{pages}{4416--4425}.
\newblock


\bibitem[Su et~al\mbox{.}(2021)]%
        {su2021noise}
\bibfield{author}{\bibinfo{person}{Wei Su}, \bibinfo{person}{Xianzhong Chen}, \bibinfo{person}{Yongguang Yu}, {and} \bibinfo{person}{Ge Chen}.} \bibinfo{year}{2021}\natexlab{}.
\newblock \showarticletitle{Noise-based control of opinion dynamics}.
\newblock \bibinfo{journal}{\emph{IEEE Trans. Automat. Control}} \bibinfo{volume}{67}, \bibinfo{number}{6} (\bibinfo{year}{2021}), \bibinfo{pages}{3134--3140}.
\newblock


\bibitem[Thorson et~al\mbox{.}(2021)]%
        {thorson2021algorithmic}
\bibfield{author}{\bibinfo{person}{Kjerstin Thorson}, \bibinfo{person}{Kelley Cotter}, \bibinfo{person}{Mel Medeiros}, {and} \bibinfo{person}{Chankyung Pak}.} \bibinfo{year}{2021}\natexlab{}.
\newblock \showarticletitle{Algorithmic inference, political interest, and exposure to news and politics on Facebook}.
\newblock \bibinfo{journal}{\emph{Information, Communication \& Society}} \bibinfo{volume}{24}, \bibinfo{number}{2} (\bibinfo{year}{2021}), \bibinfo{pages}{183--200}.
\newblock


\bibitem[T{\"o}rnberg(2022)]%
        {tornberg2022digital}
\bibfield{author}{\bibinfo{person}{Petter T{\"o}rnberg}.} \bibinfo{year}{2022}\natexlab{}.
\newblock \showarticletitle{How digital media drive affective polarization through partisan sorting}.
\newblock \bibinfo{journal}{\emph{Proceedings of the National Academy of Sciences}} \bibinfo{volume}{119}, \bibinfo{number}{42} (\bibinfo{year}{2022}), \bibinfo{pages}{e2207159119}.
\newblock


\bibitem[Wang and Shang(2015)]%
        {WANG2015180}
\bibfield{author}{\bibinfo{person}{Huanjing Wang} {and} \bibinfo{person}{Lihui Shang}.} \bibinfo{year}{2015}\natexlab{}.
\newblock \showarticletitle{Opinion dynamics in networks with common-neighbors-based connections}.
\newblock \bibinfo{journal}{\emph{Physica A: Statistical Mechanics and its Applications}}  \bibinfo{volume}{421} (\bibinfo{year}{2015}), \bibinfo{pages}{180--186}.
\newblock
\showISSN{0378-4371}


\bibitem[Wang et~al\mbox{.}(2020)]%
        {wang2020opinion}
\bibfield{author}{\bibinfo{person}{Mingwei Wang}, \bibinfo{person}{Fangshun Li}, {and} \bibinfo{person}{Decui Liang}.} \bibinfo{year}{2020}\natexlab{}.
\newblock \showarticletitle{Opinion dynamics and consensus achievement strategy based on reinforcement learning}. In \bibinfo{booktitle}{\emph{2020 IEEE International Symposium on Signal Processing and Information Technology (ISSPIT)}}. IEEE, \bibinfo{pages}{1--6}.
\newblock


\bibitem[Wongkaew et~al\mbox{.}(2015)]%
        {wongkaew2015control}
\bibfield{author}{\bibinfo{person}{Suttida Wongkaew}, \bibinfo{person}{Marco Caponigro}, {and} \bibinfo{person}{Alfio Borzi}.} \bibinfo{year}{2015}\natexlab{}.
\newblock \showarticletitle{On the control through leadership of the Hegselmann--Krause opinion formation model}.
\newblock \bibinfo{journal}{\emph{Mathematical Models and Methods in Applied Sciences}} \bibinfo{volume}{25}, \bibinfo{number}{03} (\bibinfo{year}{2015}), \bibinfo{pages}{565--585}.
\newblock


\bibitem[Xu et~al\mbox{.}(2022)]%
        {xu2022effects}
\bibfield{author}{\bibinfo{person}{Wanyue Xu}, \bibinfo{person}{Liwang Zhu}, \bibinfo{person}{Jiale Guan}, \bibinfo{person}{Zuobai Zhang}, {and} \bibinfo{person}{Zhongzhi Zhang}.} \bibinfo{year}{2022}\natexlab{}.
\newblock \showarticletitle{Effects of stubbornness on opinion dynamics}. In \bibinfo{booktitle}{\emph{Proceedings of the 31st ACM International Conference on Information \& Knowledge Management}}. \bibinfo{pages}{2321--2330}.
\newblock


\bibitem[Ye et~al\mbox{.}(2020)]%
        {ye2020continuous}
\bibfield{author}{\bibinfo{person}{Mengbin Ye}, \bibinfo{person}{Minh~Hoang Trinh}, \bibinfo{person}{Young-Hun Lim}, \bibinfo{person}{Brian~DO Anderson}, {and} \bibinfo{person}{Hyo-Sung Ahn}.} \bibinfo{year}{2020}\natexlab{}.
\newblock \showarticletitle{Continuous-time opinion dynamics on multiple interdependent topics}.
\newblock \bibinfo{journal}{\emph{Automatica}}  \bibinfo{volume}{115} (\bibinfo{year}{2020}), \bibinfo{pages}{108884}.
\newblock


\bibitem[Zhang et~al\mbox{.}(2022)]%
        {zhang2022opinion}
\bibfield{author}{\bibinfo{person}{Chengwei Zhang}, \bibinfo{person}{Dina Fang}, \bibinfo{person}{Sandip Sen}, \bibinfo{person}{Xiaohong Li}, \bibinfo{person}{Zhiyong Feng}, \bibinfo{person}{Wanli Xue}, \bibinfo{person}{Dou An}, \bibinfo{person}{Xintian Zhao}, {and} \bibinfo{person}{Rong Chen}.} \bibinfo{year}{2022}\natexlab{}.
\newblock \showarticletitle{Opinion dynamics in gossiper-media networks based on multiagent reinforcement learning}.
\newblock \bibinfo{journal}{\emph{IEEE Transactions on Network Science and Engineering}} \bibinfo{volume}{10}, \bibinfo{number}{2} (\bibinfo{year}{2022}), \bibinfo{pages}{1143--1156}.
\newblock


\bibitem[Zhang et~al\mbox{.}(2018)]%
        {DBLP:conf/icml/ZhangYL0B18}
\bibfield{author}{\bibinfo{person}{Kaiqing Zhang}, \bibinfo{person}{Zhuoran Yang}, \bibinfo{person}{Han Liu}, \bibinfo{person}{Tong Zhang}, {and} \bibinfo{person}{Tamer Basar}.} \bibinfo{year}{2018}\natexlab{}.
\newblock \showarticletitle{Fully Decentralized Multi-Agent Reinforcement Learning with Networked Agents}. In \bibinfo{booktitle}{\emph{Proceedings of the 35th International Conference on Machine Learning, {ICML} 2018, Stockholmsm{\"{a}}ssan, Stockholm, Sweden, July 10-15, 2018}} \emph{(\bibinfo{series}{Proceedings of Machine Learning Research}, Vol.~\bibinfo{volume}{80})}. \bibinfo{publisher}{{PMLR}}, \bibinfo{pages}{5867--5876}.
\newblock


\bibitem[Zhao et~al\mbox{.}(2018)]%
        {zhao2018understanding}
\bibfield{author}{\bibinfo{person}{Yiyi Zhao}, \bibinfo{person}{Gang Kou}, \bibinfo{person}{Yi Peng}, {and} \bibinfo{person}{Yang Chen}.} \bibinfo{year}{2018}\natexlab{}.
\newblock \showarticletitle{Understanding influence power of opinion leaders in e-commerce networks: An opinion dynamics theory perspective}.
\newblock \bibinfo{journal}{\emph{Information Sciences}}  \bibinfo{volume}{426} (\bibinfo{year}{2018}), \bibinfo{pages}{131--147}.
\newblock


\bibitem[Zhou et~al\mbox{.}(2023)]%
        {zhou2023sublinear}
\bibfield{author}{\bibinfo{person}{Xiaotian Zhou}, \bibinfo{person}{Liwang Zhu}, \bibinfo{person}{Wei Li}, {and} \bibinfo{person}{Zhongzhi Zhang}.} \bibinfo{year}{2023}\natexlab{}.
\newblock \showarticletitle{A sublinear time algorithm for opinion optimization in directed social networks via edge recommendation}. In \bibinfo{booktitle}{\emph{Proceedings of the 29th ACM SIGKDD Conference on Knowledge Discovery and Data Mining}}. \bibinfo{pages}{3593--3602}.
\newblock


\bibitem[Zhu et~al\mbox{.}(2021)]%
        {zhu2021minimizing}
\bibfield{author}{\bibinfo{person}{Liwang Zhu}, \bibinfo{person}{Qi Bao}, {and} \bibinfo{person}{Zhongzhi Zhang}.} \bibinfo{year}{2021}\natexlab{}.
\newblock \showarticletitle{Minimizing polarization and disagreement in social networks via link recommendation}.
\newblock \bibinfo{journal}{\emph{Advances in Neural Information Processing Systems}}  \bibinfo{volume}{34} (\bibinfo{year}{2021}), \bibinfo{pages}{2072--2084}.
\newblock


\end{thebibliography}

\end{document}